\documentclass[journal]{IEEEtran}

\usepackage{xcolor,soul,framed} 

\colorlet{shadecolor}{yellow}
\usepackage[pdftex]{graphicx}
\graphicspath{{../pdf/}{../jpeg/}}
\DeclareGraphicsExtensions{.pdf,.jpeg,.png}

\usepackage[cmex10]{amsmath}
\usepackage{bbm}
\usepackage{array}
\usepackage{eqparbox}
\usepackage{url}

\usepackage{import}
\usepackage{algorithm}
\usepackage{algorithmic}
\usepackage{mathtools}
\usepackage{amsthm}
\usepackage{amsfonts}
\usepackage{tikz}
\usepackage{mathdots}
\usepackage{yhmath}
\usepackage{cancel}
\usepackage{color}
\usepackage{siunitx}
\usepackage{array}
\usepackage{multirow}
\usepackage{amssymb}
\usetikzlibrary{fadings}
\DeclareMathOperator*{\argmax}{arg\,max}
\DeclareMathOperator*{\argmin}{arg\,min}

\newtheorem{theorem}{Theorem}[section]
\newtheorem{corollary}{Corollary}[theorem]

\newcommand\ddfrac[2]{\frac{\displaystyle #1}{\displaystyle #2}}

\makeatletter
\def\thm@space@setup{%
  \thm@preskip=5pt
  \thm@postskip=\thm@preskip 
}
\makeatother

\begin{document}
    \title{Maximal Information Leakage based Privacy Preserving Data Disclosure Mechanisms}
  \author{Tianrui~Xiao,~\IEEEmembership{Student Member,~IEEE,}
          and~Ashish~Khisti,~\IEEEmembership{Member,~IEEE}
\thanks{$^{1}$ T. Xiao is with Faculty of Electrical \& Computer Engineering,
        University of Toronto, 10 King's College Road, Toronto, Ontario Canada M5S 3G4
        {\tt\small tianrui.xiao at mail.utoronto.ca}}%
\thanks{$^{2}$ A. Khisti is with the Faculty of Electrical \& Computer Engineering,
        University of Toronto, 10 King's College Road, Toronto, Ontario Canada M5S 3G4
        {\tt\small akhisti at ece.utoronto.ca}}%
}


\maketitle

\begin{abstract}
It is often necessary to disclose training data to the public domain, while protecting privacy of certain sensitive labels.  We use information theoretic measures to develop such privacy preserving data disclosure mechanisms. Our  mechanism involves perturbing the data vectors in a manner that strikes a balance in the privacy-utility trade-off.
We use  maximal information leakage between the output data vector and the confidential label as our privacy metric. We first study the theoretical Bernoulli-Gaussian model and study the privacy-utility trade-off when only the mean of the Gaussian distributions 
can be perturbed. We show that the optimal solution is the same as the case when the utility is measured using  probability of error at the adversary. We then consider an application of this framework to a data driven setting and provide an empirical approximation to the Sibson mutual information.  By performing experiments on the MNIST and FERG data-sets, we show that our proposed framework achieves equivalent or better privacy than previous methods based on mutual information.
\end{abstract}


%
\IEEEpeerreviewmaketitle


\section{Introduction}
\par In the area of data disclosure and information privacy, one of the fundamental questions of interest is how much information is leaked when an observation is made about a correlated quantity. The observation is considered to be information provided to a (possibly malignant) adversary, and it is in our interest to protect the sensitive information. While disclosure of information to an adversary may be intentional, such as publishing statistical information regarding a data set, in many scenarios this is unintentional, and may lead to security breaches or leakage of sensitive information. The focus of this paper is to address the problem of applying transformations to sensitive data for disclosure while protecting privacy using an information theoretic framework.
\par In the broader literature, privacy preserving data disclosure is a widely explored area motivated by highly publicized data breaches which resulted from inadequate anonymization techniques \cite{SweeneySimpleDemographics} \cite{NarayananRobustDeanon}. Many methods have been proposed to statistically quantify and measure privacy, including k-anonymity, t-closeness, Arimoto mutual information of order $\infty$\cite{DBLP:conf/fossacs/Smith09}, $\max_{P_X}I_{\infty}(X;Z)$ \cite{DBLP:journals/entcs/BraunCP09} (\cite{DBLP:conf/csfw/BartheK11} studies the same metric in a differential privacy context) and more recently mutual information \cite{PPANpaper}\cite{MinimaxFilter}.  Work has been done in the area of differential privacy\cite{DworkFoundationofDiffPriv} utilizing data-driven frameworks developed in deep learning\cite{DeepLearningwDiffPriv}, in particular private machine learning through noisy stochastic gradient descent(SGD) or private aggregation of teacher ensembles(PATE) (\cite{DifPrivEmpRiskMin, SGDwithPrivUpdates, PrivfromMultipartyData, Shokri:2015:PDL:2810103.2813687}). Prior work also borrow from the information theory literature to design machine learning models to achieve domain-specific goals such as exploration in reinforcement learning \cite{VIME}.
\par Numerous adversarial learning techniques have been proposed in recent years, spearheaded by the development of generative adversarial networks(GAN) and subsequent variants \cite{GoodfellowGAN} \cite{ImprovedtechniquesGANs}. Under the GAN framework, the model is composed of a discriminator and a generator, where the discriminator's objective is to classify whether or not input samples are real or generated, and the generator's objective is to produce samples that fool the discriminator. There have been different variations on conditioning for the input in order to learn more flexible spaces and provide interpretation of the input space for the generator, as well as learning representations for specific types of data (\cite{InfoGAN} \cite{CycleGAN} \cite{MakhzaniAdvAutoencoders}). 
\par Previous works predominantly adopt classic information-theoretic measures like Shannon-entropy and mutual information to quantify the amount of information leaked between the disclosed variable and the private variable~\cite{PPANpaper}\cite{MinimaxFilter}. The main advantage of using an information theoretic measure of privacy is that it considers the statistical distribution of the data. The authors use a min-max formulation of an generative adversarial network to achieve a trade-off between distortion and concealing private information by means of a randomized function implemented as a neural network. A similar approach was adopted by Huang et al\cite{CGAPpaper} in which the authors consider two losses for a similar adversarial model, the 0-1 loss and the empirical log-loss, each corresponding to the maximum a posteriori (MAP) adversary and the minimum cross-entropy adversary. Their notion of using the probability of a correct guess of an adversary as the metric was first studied in \cite{DBLP:journals/information/AsoodehDAL16} \cite{DBLP:conf/cwit/AsoodehAL15}. The log-loss in the model from \cite{CGAPpaper} was shown to approach the game-theoretic optimal mechanisms under a MAP adversary, and it also recovers mutual information privacy.
\par Maximal information leakage is motivated by a guessing adversary to characterize the amount of information the public variable $Z$ leaks about a confidential variable $C$. Leakage is defined as the logarithm of the ratio of an adversary's probability of a correct guess of a (randomized) function of $C$ denoted as $\hat{U}(C)$ when $Z$ is observed, to the probability of a correct blind guess. The maximal information leakage then is defined as the maximum leakage over all possible functions. Since the leakage is maximized over the random variable $U$ with the Markov chain $U-C-Z$, it represents the worst case of possible functions of $U$. In \cite{IssaWagnerOperationalMeasure} the maximization is proven to admit a closed-form solution and is proven to be equal to the Sibson mutual information of order infinity. We note that prior works on maximal information leakage also include~\cite{AlvimGeneralizedGainFunctions,DBLP:conf/fossacs/Smith09,Axiomsforleakage}.
\section{Preliminaries: Sibson Mutual Information and Information Leakage}
\par Here we formally introduce the concepts of Sibson mutual information and maximal information leakage. R\'enyi introduced generalized definitions of Shannon entropy and KL divergence in R\'enyi entropy and R\'enyi divergence (equation (\ref{eq:Renyi divergence def})) which later was used in lossless data compression\cite{Csiszar95GenCutoffandRenyiInfoMeasures} and hypothesis testing\cite{Ben-BassatR78RenyiEntropy}. However, he did not generalize mutual information, and several approaches have been proposed in the literature\cite{VerdualphaMI}. Sibson mutual information is an information theoretic measure based on a generalization of mutual information, defined in equation (\ref{eq:SibMIoriginal def}) for random variables $X \in \mathcal{X}$, $Y \in \mathcal{Y}$ distributed as $P(X,Y)$.
\begin{flalign}
&I_{\alpha}(X;Y) = \min_{Q_Y}D_{\alpha}(P_{Y|X}||Q_Y|P_X) \label{eq:SibMIoriginal def}\\
&D_{\alpha}(P||Q) = \frac{1}{\alpha - 1}\log\Big(\displaystyle\sum_{a \in \mathcal{A}}P^{\alpha}(a)Q^{1 - \alpha}(a)\Big) \label{eq:Renyi divergence def}
\end{flalign}
For discrete variables, the Sibson mutual information is
\begin{flalign}
&I_{\alpha}(X;Y) = \frac{\alpha}{\alpha - 1} \log \displaystyle\sum_{y \in \mathcal{Y}}\Big(\displaystyle\sum_{x \in \mathcal{X}} P_X(x)P_{Y|X=x}^{\alpha}(y)\Big)^{1/\alpha}
\end{flalign}
This definition of Sibson mutual information in the limit as $\alpha \rightarrow \infty$
is shown to be equal to the maximal information leakage\cite{IssaW17OperationalDefinitions}
\begin{flalign}
&\mathcal{L}(X \rightarrow Y) = \displaystyle\sup_{U-X-Y-\hat{U}} \log \frac{Pr(U=\hat{U})}{\max_{u \in \mathcal{U}}P_U(u)}\\
&= \log \displaystyle\sum_{y \in \mathcal{Y}} \displaystyle\max_{x \in \mathcal{X}: P_X(x) > 0} P_{Y|X}(y|x) = I_{\infty}(X;Y)
\end{flalign}
\par Operationally, the information leakage is considered as the logarithm of the multiplicative increase in an adversary's ability to predict $U$, a (randomized) function of $X$ in $\hat{U}$, having observed $Y$ compared to a blind guess(\cite{IssaWagnerOperationalMeasure, IssaW17OperationalDefinitions}). The maximal information leakage, then, is the maximization of the leakage over all such randomized functions $U$. This is a conservative measure, and it has certain desirable properties that are demonstrated in (\cite{IssaW17OperationalDefinitions, IssaWagnerOperationalMeasure, VerdualphaMI}).
\par While mutual information is widely used (as exemplified in related work \cite{PPANpaper} \cite{CGAPpaper}), there are many scenarios where it is unable to capture the performance of a MAP adversary for a given mapping, as the example below demonstrates.
Consider a $C$ variable as a 2k-bit integer distributed as a uniform distribution over the possible $2^{2k}$ values ($k \ge 2)$, and the following two mappings:
\begin{align*}
    Z_1 = \begin{cases} C, \quad C\; mod\; 2 = 0\\
                        1, \quad else
        \end{cases}
\end{align*}
\begin{align*}
    Z_2 = C \& (0^{k-1}1^{k+1})
\end{align*}
where $Z_1$ is preserved to be $C$ if the last bit in $C$ is 0, and $Z_2$ is the mapping which preserves the last $k+1$ bits of $C$ as the logical AND operator zeros out the first $k-1$ bits. Under these mappings, one can easily compute the mutual information as follows:\\
\begin{align*}
    I(C;Z_1) = \frac{1}{2}\log(\frac{2}{1}) + 2^{2k-1} * 2^{-2k} \log(2^{2k}) = k + \frac{1}{2}\\
    I(C;Z_2) = k+1
\end{align*}
Note that the mutual information in the two mapping is nearly identical. In terms of an adversary's performance, a MAP adversary can correctly guess $C$, $1/2$ of the time in the first mapping, whereas the second mapping has an MAP adversary accuracy of $\frac{1}{2^{k-1}}$. When calculating the maximal information leakage for these two mappings (c.f.\ example 3 of \cite{IssaWagnerOperationalMeasure})  yields:
\begin{flalign}
    &I_\infty(C;Z_1) = \log(|\{z; P_Z(z) > 0\}|) \\
    &= \log(2^{2k-1} + 1) \approx 2k-1 \nonumber\\
    &I_\infty(C;Z_2) = \log(2^{k+1}) = k+1
\end{flalign}
Then it is clear that the maximal leakage in the first mapping is nearly twice that of the second mapping, which is consistent with the fact that an adversary can guess $C$ based on $Z_1$ better than based on $Z_2$.


\section{Contributions}
Previous approaches (\cite{PPANpaper}, \cite{CGAPpaper}) used conventional mutual information as a metric to derive privatizer-adversary models for theoretical Gaussian data and the MNIST data set. We study the utility of using maximal information leakage as a privacy measure in this paper. 

In section \ref{sec: theoretical gaussian section} we introduce an optimization problem for affine transformations on Gaussian data, and show solutions for this optimization problem, which are extended based on the work in \cite{CGAPpaper}. We then consider three different objectives as our privacy metric (1) the MAP adversary accuracy (2) Maximal information leakage and (3) an approximation of Sibson mutual information; Interestingly all three metrics are then shown to result in the same optimization problem and thus identical affine transformation can be used regardless of the metric. We also briefly consider an extension of the transformation with noise, and show that global optimum are not known analytically. 

In section \ref{sec: data driven approach} we adapt our setup to be used in models where we have access to data samples drawn from the distribution without knowing the parameters of the distribution. Section \ref{sec: experiments section} demonstrates results from synthetic Gaussian data where we can compare with theoretical MAP adversary accuracies, the MNIST data set, and FERG data set, and we conclude in section \ref{sec: conclusion section}.
We propose to use an GAN-like setup where we simultaneously train two models: (1) an adversarial classification model which has access to the training set along with private labels and (2) an auto-encoder to implement a randomized privatizer that is subjected to a distortion constraint and a privacy constraint using Sibson mutual information. By carefully training both the models in tandem we show that significant improvements can be attained in the privacy-utility trade-off. For the FERG data set, we design a variant of the auto-encoding model to measure the utility based on the adversary's ability to infer a related public variable rather than just the reconstruction.

\section{Affine Transformations of Gaussian Data} \label{sec: theoretical gaussian section}
In this section we use a Gaussian data setting and affine transformations with a distortion budget identical to the setup used in \cite{CGAPpaper} to define an optimization problem (equation (\ref{eq:initial optimization problem def})) that is aimed to preserve privacy. This data setting is chosen since the Gaussian distribution is ubiquitous in many applications \cite{NormalDistributions}. Affine transformations preserve Gaussianity of the data, allowing the problem to be more tractable, and in a later extension we consider a noisy transformation. We then show that there are two solutions conditional on the distortion budget, one of which is same as the result given by \cite{CGAPpaper} in their game-theoretic solution to the optimization when using MAP adversary accuracy as the optimization objective. Starting from MAP adversary accuracy as the objective function, we demonstrate that it is equivalent to the optimization problem of equation (\ref{eq:initial optimization problem def}), and hence there are two solutions instead of the one proposed in \cite{CGAPpaper}. We then consider the maximal information leakage as the objective, and reduce the optimization to that of equation (\ref{eq:initial optimization problem def}), thus demonstrating that its solutions are identical to that of equation (\ref{eq:initial optimization problem def}). We also consider Sibson mutual information as an objective, and demonstrate that with a numerical approximation, its optimization is again equal to the optimization in equation (\ref{eq:initial optimization problem def}), yielding the same solutions. We finally consider a noisy transformation and demonstrate that the optimization of Sibson mutual information for this transformation does not guarantee an analytic global solution, same as prior work \cite{CGAPpaper} did with MAP adversary accuracy as the metric for the same class of transformations.

\subsection{Gaussian data definitions}
This is a theoretical data setting where the privatizer controlling the transform and the adversary inferring a private variable both have access to the joint distributions of the public variable $X$ and the private variable $C$ as $P(X,C)$. $X$ follows a mixture of Gaussian distribution:
\begin{multline}\label{eq:1}
    p(X|C=0) \sim \mathcal{N}(\mu_0, \sigma^2), \quad p(X|C=1) \sim \mathcal{N}(\mu_1, \sigma^2)
\end{multline}
with conditional probabilities 
\begin{flalign}\label{eq:2}
    P(C=0) = \tilde{p}, \quad P(C=1) = 1-\tilde{p}
\end{flalign}
W.L.O.G. we may let $\mu_0 \le \mu_1$. The Gaussian distributions have equal covariance for tractability purposes.

\subsection{Affine transformation} \label{sec: affine transform def}
\begin{figure}[!h]
\tikzset{every picture/.style={line width=0.75pt}} 
\resizebox {\columnwidth} {!} {
\begin{tikzpicture}[x=0.75pt,y=0.75pt,yscale=-1,xscale=1]

\draw  (17,125) -- (435,125)(226.72,29) -- (226.72,146) (428,120) -- (435,125) -- (428,130) (221.72,36) -- (226.72,29) -- (231.72,36)  ;
\draw [color={rgb, 255:red, 134; green, 206; blue, 63 }  ,draw opacity=1 ]   (273,118) .. controls (314,110) and (330,31) .. (358,63) .. controls (386,95) and (389,115) .. (419,120) ;

\draw [color={rgb, 255:red, 17; green, 22; blue, 231 }  ,draw opacity=1 ]   (34,116) .. controls (75,108) and (91,29) .. (119,61) .. controls (147,93) and (151,115) .. (179,116) ;

\draw [color={rgb, 255:red, 17; green, 22; blue, 231 }  ,draw opacity=1 ]   (102,116) .. controls (143,108) and (159,29) .. (187,61) .. controls (215,93) and (219,115) .. (247,116) ;

\draw [color={rgb, 255:red, 134; green, 206; blue, 63 }  ,draw opacity=1 ]   (228,117) .. controls (269,109) and (285,30) .. (313,62) .. controls (341,94) and (344,114) .. (374,119) ;

\draw    (107,53) -- (169,52.03) ;
\draw [shift={(171,52)}, rotate = 539.1] [color={rgb, 255:red, 0; green, 0; blue, 0 }  ][line width=0.75]    (10.93,-3.29) .. controls (6.95,-1.4) and (3.31,-0.3) .. (0,0) .. controls (3.31,0.3) and (6.95,1.4) .. (10.93,3.29)   ;

\draw    (345,54) -- (304,54) ;
\draw [shift={(302,54)}, rotate = 360] [color={rgb, 255:red, 0; green, 0; blue, 0 }  ][line width=0.75]    (10.93,-3.29) .. controls (6.95,-1.4) and (3.31,-0.3) .. (0,0) .. controls (3.31,0.3) and (6.95,1.4) .. (10.93,3.29)   ;

\draw (99,142) node   {$P( X|C=0)$};
\draw (359,141) node   {$P( X|C=1)$};
\draw (138,41) node   {$\beta _{0}$};
\draw (324,41) node   {$\beta _{1}$};
\draw (439,109) node   {$x$};
\draw (229,14) node   {$P( x)$};

\end{tikzpicture}
}
\caption{Binary Gaussian data and transformation vectors}
\label{fig:Affine transform}
\end{figure}
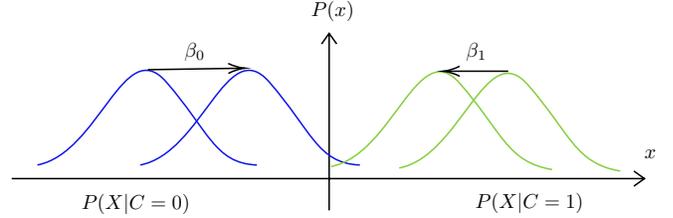

We define the following data-dependent affine transformation:
\begin{flalign} \label{eq: affine transform}
&Z = X + (1 - C)\beta_0 - C\beta_1
\end{flalign}
This transformation is dependent on the parameters $\beta_0, \beta_1$, and can be seen in Figure \ref{fig:Affine transform}.
\begin{flalign}\label{eq:3}
    p(Z|C=0) \sim \mathcal{N}(\mu_0 + \beta_0, \sigma^2) = \mathcal{N}(\mu_0^{'}, \sigma^2), \\
    p(Z|C=1) \sim \mathcal{N}(\mu_1 - \beta_1, \sigma^2) = \mathcal{N}(\mu_1^{'}, \sigma^2) \\
    \beta_0, \beta_1 \ge 0 \\
    \mu_0^{'} \le \mu_1^{'} \label{eq:6}
\end{flalign}
The $Z$ distribution conditioned on the class $C$ are defined by its means $\mu_0^{'}, \mu_1^{'}$ and variance $\sigma^2$. The adversary knows the distribution of $Z$ and therefore only needs to compute its guess via the MAP decision rule given $Z$.

\subsection{Optimization problem and solutions}\label{sec: affine optimization problem def}
With the affine transformation, we define an additional distortion constraint based on a distortion budget denoted as $D$ as a measure of utility:
\begin{flalign}
    \mathcal{D} = \{(\beta_0, \beta_1)|(1-\tilde{p})\beta_0^2 + \tilde{p}\beta_1^2 \le D , \beta_0 \ge 0, \beta_1 \ge 0\} \label{eq:affine D definition}
\end{flalign}
Under the aforementioned transformations we consider the following optimization problem:
\begin{flalign}
&\max_{(\beta_0, \beta_1) \in \mathcal{D}} \frac{\mu_0^{'} - \mu_1^{'}}{2\sigma}. \label{eq:initial optimization problem def}
\end{flalign}
The solution to this optimization problem is 
\begin{equation}
    \beta_0^* = \sqrt{\frac{\tilde{p}}{1 - \tilde{p}} D}, \qquad \beta_1^* = \sqrt{\frac{1 - \tilde{p}}{\tilde{p}} D} \label{eq:affine solution 1}
\end{equation}
if $D$ satisfies
\begin{flalign}
&D \le \tilde{p}(1 - \tilde{p})(\mu_1 - \mu_0)^2 \label{eq:affine solution condition}
\end{flalign}
and
\begin{flalign}
&\beta_0^* = (\mu_1 - \mu_0)(1 - \tilde{p}), \\
&\beta_1^* = (\mu_1 - \mu_0)\tilde{p} \label{eq:affine solution 2}\nonumber
\end{flalign}
otherwise.
Refer to Appendix 1 Section A for detailed solutions.

In the following subsections we will consider optimizing over the transformation specified in \ref{sec: affine transform def} with three different privacy metrics as the objective function: MAP adversary accuracy, maximal information leakage, and Sibson mutual information. Interestingly we will show that all three optimization problems are related to~\eqref{eq:initial optimization problem def} and the solution in this section gives the parameters of the optimal transformation. 

\subsection{MAP accuracy as a metric}
In this section we consider the optimization for the transformations in section \ref{sec: affine transform def} with the MAP adversary's accuracy as the privacy metric, as prior work \cite{CGAPpaper} has done. Their theorem provides the solution in equation (\ref{eq:affine solution 1}) but not the solution in equation \eqref{eq:affine solution 2} when condition \eqref{eq:affine solution condition} is not satisfied. The optimization problem is
\begin{flalign}
    \min_{(\beta_0, \beta_1) \in \mathcal{D}} Pr(\hat{C} = C)
\end{flalign}
where $Pr(\hat{C} = C)$ is the MAP adversary's accuracy. We can characterize the adversary's accuracy in terms of the distortion constraint and the optimal transformation with the following theorem:
\begin{theorem} \label{thm:MAPacc affine}
Under the binary Gaussian data scenario with affine transformations of the data described in the set of equations and inequalities (\ref{eq:1}) - (\ref{eq:6}) over the set $\mathcal{D}$, the adversary's accuracy after solving the optimization for the optimal parameters $(\beta_0^*, \beta_1^*)$
\begin{flalign}
    (\beta_0^*, \beta_1^*) = \argmin_{(\beta_0, \beta_1) \in \mathcal{D}} Pr(\hat{C} = C)
\end{flalign} is
\begin{flalign}
&Pr^*(\hat{C} = C) = \tilde{p}Q\big(\frac{\sigma}{\mu_0^{'} - \mu_1^{'}}\log(\frac{1 - \tilde{p}}{\tilde{p}}) - \frac{\mu_0^{'} - \mu_1^{'}}{2 \sigma}\big) + \\
&(1 - \tilde{p})Q\big(- \frac{\sigma}{\mu_0^{'} - \mu_1^{'}}\log(\frac{1 - \tilde{p}}{\tilde{p}}) - \frac{\mu_0^{'} - \mu_1^{'}}{2 \sigma}\big) 
\label{eq: MAP accuracy expression}
\end{flalign}
where the $Q(\cdot)$ function is 
\begin{flalign}
&Q(x) = \frac{1}{\sqrt{2 \pi}} \int_x^{\infty} e^{-\frac{u^2}{2}} du
\end{flalign}
and the solutions $\beta_0^*, \beta_1^*$ are given by equations (\ref{eq:affine solution 1}), (\ref{eq:affine solution 2}).\\
\textbf{Proof:} Refer to Appendix 1 Section B.
\end{theorem}
Note that the above solution is under the assumption that $\mu_0 \le \mu_1$ and $\mu_0^{'} \le \mu_1^{'}$, since the MAP decision rule would be reversed if the means are shifted over each other. In \cite{CGAPpaper}, their game theoretic solutions are the same as ours for optimization over the MAP adversary accuracy in equation (\ref{eq:affine solution 1}), but we specify a constraint on the distortion budget $D$ (equation (\ref{eq:affine solution condition})) that gives another solution (equation (\ref{eq:affine solution 2})) when the condition is not satisfied.

\subsection{Maximal Information Leakage as a metric}
Now we propose using maximal information leakage as an optimization metric, and investigate the induced optimization problem based on the same synthetic data distributions and affine transformation as the previous section. The optimization solution is now given by:
\begin{flalign}
    (\beta_0^*, \beta_1^*) = \argmin_{(\beta_0, \beta_1) \in \mathcal{D}} I_{\infty}(C; Z)
\end{flalign}
The following theorem relates the optimization problem to the optimization in equation (\ref{eq:initial optimization problem def}), and characterizes the solutions of the optimization.
\begin{theorem} \label{thm:MaxIL affine}
Under binary mixture of Gaussians data described in equations \eqref{eq:1} - \eqref{eq:6} over the set $\mathcal{D}$, assuming $\mu_0^{'} < \mu_1^{'}$, the solution to minimization of maximal information leakage is equal to
\begin{flalign}
&(\beta_0^*, \beta_1^*) = \argmin_{(\beta_0, \beta_1) \in \mathcal{D}} \log\big(2 Q(\frac{\mu_0^{'} - \mu_1^{'}}{2\sigma})\big) = \\
&\argmax_{(\beta_0, \beta_1) \in \mathcal{D}} \big(\frac{\mu_0^{'} - \mu_1^{'}}{2\sigma}\big)
\end{flalign}
and the solutions $\beta_0^*, \beta_1^*$ are given by equations (\ref{eq:affine solution 1}, \ref{eq:affine solution 2}).\\
\textbf{Proof:} Under the mixture of Gaussians distribution and assuming that $\mu_0^{'} < \mu_1^{'}$, we have:
\begin{equation}
    I_{\infty}(C; Z) = \log \Big(\displaystyle\int_{-\infty}^{z_0} p_{Z|C=0} + \displaystyle\int_{z_0}^{\infty} p_{Z|C=1}\Big)
\end{equation}
The intersection point can be found in this scenario as 
\begin{equation}
    z_0 = \frac{\mu_1^{'2} - \mu_0^{'2}}{2(\mu_1^{'} - \mu_0^{'})} = \frac{\mu_1^{'} + \mu_0^{'}}{2}
\end{equation}
Hence solving the optimization objective of minimizing the maximal information leakage subject to a distortion constraint is equivalent to:
\begingroup
\allowdisplaybreaks
\begin{flalign}
&\argmin_{(\beta_0, \beta_1) \in \mathcal{D}} \log\Big((1-Q(\frac{z_0 - \mu_0^{'}}{\sigma})) + Q(\frac{z_0 - \mu_1^{'}}{\sigma})\Big) \\
&= \argmin_{(\beta_0, \beta_1) \in \mathcal{D}} \log\Big((1-Q(\frac{\mu_1^{'} - \mu_0^{'}}{2\sigma})) + Q(\frac{\mu_0^{'} - \mu_1^{'}}{2\sigma})\Big) \\
&= \argmin_{(\beta_0, \beta_1) \in \mathcal{D}} \log\Big(2 Q(\frac{\mu_0^{'} - \mu_1^{'}}{2\sigma})\Big) = \argmax_{(\beta_0, \beta_1) \in \mathcal{D}} \frac{\mu_0^{'} - \mu_1^{'}}{2\sigma}
\end{flalign}
\endgroup
The optimization is the same as the one proposed in equation (\ref{eq:initial optimization problem def}), subject to the constraints specified in equation (\ref{eq:6}) and (\ref{eq:affine D definition}), and yields the same results for $\beta_0^*, \beta_1^*$
\qedsymbol
\end{theorem}
Therefore when optimizing the maximal information leakage for the defined data distribution and transformation, it is equivalent to minimizing an adversary's theoretical performance, and both reduce to minimizing the normalized distance between the means of the transformed Gaussian distributions.

\subsection{Sibson mutual information as a metric} \label{sec:affine sibmi sec}
Here we consider affine transformations of data distributed as a mixture of Gaussians conditioned on their class specified in equations \eqref{eq:1}-\eqref{eq:6}, with Sibson mutual information as the privacy metric in the optimization. Since the maximal information leakage is equal to the Sibson mutual information of order $\infty$ \cite{IssaWagnerOperationalMeasure}, we will approximate it with Sibson mutual information of order $\alpha$. The goal is to solve the following optimization problem with respect to the parameters $\beta_0, \beta_1$:
\begin{flalign*}
    (\beta_0^*, \beta_1^*) = \argmin_{(\beta_0, \beta_1) \in \mathcal{D}} I_{\alpha}(C; Z)
\end{flalign*}
The following theorem relates the optimization of Sibson mutual information to the optimization in equation \eqref{eq:initial optimization problem def} and characterizes the solutions.
\begin{theorem} \label{thm:SibMI affine}
Under binary mixture of Gaussians data described by equations \eqref{eq:1} - \eqref{eq:6} over the set $\mathcal{D}$, the solution to the minimization of Sibson mutual information is equal to
\begin{flalign}
&(\beta_0^*, \beta_1^*) = \argmin_{(\beta_0, \beta_1) \in \mathcal{D}} I_{\alpha}(C; Z) \approx \argmax_{(\beta_0, \beta_1) \in \mathcal{D}} \frac{\mu_0^{'} - \mu_1^{'}}{\sigma}
\end{flalign}

and the approximate solutions $\beta_0^*, \beta_1^*$ are given by equations (\ref{eq:affine solution 1}, \ref{eq:affine solution 2}).\\

\textbf{Proof:} Based on the definition of Sibson mutual information we have:
\small
\begin{flalign}
&I_\alpha(C; Z) = \frac{\alpha}{\alpha - 1} \log\Big(\displaystyle\int_z \sum_c(P_{Z|C}^{\alpha}(z|c)P_C(c))^{1/\alpha} dz\Big)\\
&= \frac{\alpha}{\alpha - 1} \log\Big(\displaystyle\int_z (P_{Z|C=0}^{\alpha}P_{C=0} + P_{Z|C=1}^{\alpha}(1-P_{C=0}))^{1/\alpha} dz\Big) \\
&= \frac{\alpha}{\alpha - 1} \log\Big( \displaystyle\int_z P_{Z|C=0}P_{C=0}^{1/\alpha} (1 + \frac{1 - P_{C=0}}{P_{C=0}} \frac{P_{Z|C=1}^{\alpha}}{P_{Z|C=0}^{\alpha}})^{1/\alpha} dz\Big) \\
&\approx \frac{\alpha}{\alpha - 1} \log\Big( \displaystyle\int_z P_{Z|C=0}\tilde{p}^{1/\alpha} \max(1, (\frac{1 - \tilde{p}}{\tilde{p}})^{1/\alpha} \frac{P_{Z|C=1}}{P_{Z|C=0}}) dz\Big) \label{eq:sibMIapproximation}\\
&= \frac{\alpha}{\alpha - 1} \log\Big( \displaystyle\int_{-\infty}^{z_0} \tilde{p}^{1/\alpha}P_{Z|C=0} dz + \displaystyle\int_{z_0}^{\infty}(1 - \tilde{p})^{1/\alpha} P_{Z|C=1} dz\Big)
\end{flalign}
\begin{flalign}
&z_0 = \ddfrac{\frac{2\sigma^2}{\alpha}\log(\frac{1-\tilde{p}}{\tilde{p}}) + \mu_0^{'2} - \mu_1^{'2}}{2 (\mu'_0 - \mu'_1)}, \quad \mu_0^{'} \le \mu_1^{'}
\end{flalign}
\normalsize
\par We approximate the inner term with a max function, allowing us to express the integral in a piece-wise fashion. This approximation in numerical simulations was sufficiently close $(99.8\%)$ to the true value of the Sibson mutual information of the same order for the case of binary Gaussian data on orders of $20$ or greater. The $z_0$ derived under this metric is equivalent to the one derived from maximal information leakage for high orders of $\alpha$, and the resulting optimization is cast as
\small
\begin{flalign}
&(\beta_0^*, \beta_1^*) = \argmin_{(\beta_0, \beta_1) \in \mathcal{D}} \frac{\alpha}{\alpha - 1} \log(\tilde{p}^{1/\alpha} Q(-\frac{z_0 - \mu_0^{'}}{\sigma}) + \\
&(1 - \tilde{p})^{1/\alpha} Q(\frac{z_0 - \mu_1^{'}}{\sigma})) \nonumber\\
&= \argmin_{(\beta_0, \beta_1) \in \mathcal{D}} \frac{\alpha}{\alpha - 1} \log(\tilde{p}^{1/\alpha} Q(-\frac{\frac{\sigma}{\alpha}\log(\frac{1 - \tilde{p}}{\tilde{p}})}{\mu_0^{'} - \mu_1^{'}} + \frac{\mu_0^{'} - \mu_1^{'}}{2\sigma}) + \\
&(1 - \tilde{p})^{1/\alpha} Q(\frac{\frac{\sigma}{\alpha}\log(\frac{1 - \tilde{p}}{\tilde{p}})}{\mu_0^{'} - \mu_1^{'}} + \frac{\mu_0^{'} - \mu_1^{'}}{2\sigma})) \nonumber\\
&= \argmin_{(\beta_0, \beta_1) \in \mathcal{D}} \frac{\alpha}{\alpha - 1} \log(\tilde{p}^{1/\alpha}Q(\frac{1}{d \alpha}\log(\frac{1 - \tilde{p}}{\tilde{p}}) - \frac{d}{2}) + \\
&(1 - \tilde{p})Q(- \frac{1}{d \alpha}\log(\frac{1 - \tilde{p}}{\tilde{p}}) - \frac{d}{2})), \quad d = \frac{\mu_1^{'} - \mu_0^{'}}{\sigma} \nonumber\\
&= \argmin_{(\beta_0, \beta_1) \in \mathcal{D}} d = \argmax_{(\beta_0, \beta_1) \in \mathcal{D}} \frac{\mu_0^{'} - \mu_1^{'}}{\sigma} \label{eq: SibMI opt equals argmax d}
\end{flalign}
\normalsize
Equation (\ref{eq: SibMI opt equals argmax d}) is derived in the same way as Appendix 1.B and is shown in Appendix 1.C. Note that this is the same optimization as equation (\ref{eq:initial optimization problem def}) with the same constraints specified in equation (\ref{eq:6}) and (\ref{eq:affine D definition}), so the optimization will recover the same solution.
\qedsymbol
\end{theorem}
Under the approximation for Sibson mutual information, we show that in the limit as $\alpha$ approaches $\infty$, the approximation approaches the definition for maximal information leakage.

\begin{flalign}
& z_0 = \ddfrac{\frac{2\sigma^2}{\alpha}\log(\frac{1-\tilde{p}}{\tilde{p}}) + \mu_0^{'2} - \mu_1^{'2}}{2 (\mu'_0 - \mu'_1)}, \quad \mu_0^{'} \le \mu_1^{'}\\
&\displaystyle\lim_{\alpha \rightarrow \infty} I_{\alpha}(C;Z) \approx \displaystyle\lim_{\alpha \rightarrow \infty} \frac{\alpha}{\alpha - 1} \log\Big( \displaystyle\int_{-\infty}^{z_0} \tilde{p}^{1/\alpha}P_{Z|C=0} dz\\
&+ \displaystyle\int_{z_0}^{\infty}(1 - \tilde{p})^{1/\alpha} P_{Z|C=1} dz\Big) \nonumber\\
&= \log\Big( \displaystyle\int_{-\infty}^{z_0'} P_{Z|C=0} dz + \displaystyle\int_{z_0'}^{\infty} P_{Z|C=1} dz\Big)\\
& z_0' = \ddfrac{\mu_0^{'2} - \mu_1^{'2}}{2 (\mu'_0 - \mu'_1)}, \quad \mu_0^{'} \le \mu_1^{'}
\end{flalign}
From theorem (\ref{thm:MAPacc affine}-\ref{thm:SibMI affine}) we can infer the following corollary:
\begin{corollary} \label{cor: corollary connecting the solutions for affine transformation}
Under the binary mixture of Gaussian data and affine transformations given by equations (\ref{eq:1}) - (\ref{eq:6}), the solutions to optimization over the adversary performance, maximal information leakage, and Sibson mutual information approximation are the same.
\begin{flalign}
    \argmin_{(\beta_0, \beta_1) \in \mathcal{D}} I_\alpha(C; Z) \approx \argmin_{(\beta_0, \beta_1) \in \mathcal{D}} I_\infty(C; Z) \\
    = \argmin_{(\beta_0, \beta_1) \in \mathcal{D}} Pr(\hat{C}=C)
\end{flalign}
\end{corollary}

\subsection{Extension to transformations with class-independent noise} \label{sec:Affine noisy extension}
We now consider a class of transformations with the same initial binary mixture of Gaussian data described in equations (\ref{eq:1}) - (\ref{eq:2}), but with the following transformation:
\begin{flalign}
    Z = X + (1 - C)\beta_0 - C\beta_1 + \gamma N \label{eq:affine with noise}\\
    N \sim \mathcal{N}(0, 1)
\end{flalign}
This is an affine transformation with added Gaussian noise, which preserves Gaussianity of the $Z$ distribution, and still maintains tractability for analyzing the optimization problem.
Our distortion constraint is adjusted to account for the independent noise and is defined as
\begin{flalign}
    (1 - \tilde{p})\beta_0^2 + \tilde{p}\beta_1^2 + \gamma^2 \le D\\
    \beta_0, \beta_1, \gamma \ge 0
\end{flalign}
Thus our optimization problem is
\begin{flalign}
    \min I_{\alpha}(C;Z)\\
    s.t.\quad (1 - \tilde{p})\beta_0^2 + \tilde{p}\beta_1^2 + \gamma^2 \le D, \\
    \beta_0, \beta_1, \gamma \ge 0
\end{flalign}
\begin{theorem}
    For the data over $X, C$ described in equations (\ref{eq:1}), (\ref{eq:2}), and the data transformation in equation (\ref{eq:affine with noise}), the optimal parameters $\beta_0^*, \beta_1^*, \gamma^*$ are given as the solution to
    \begin{flalign}
&min_{\beta_0, \beta_1, \gamma} \frac{(\mu_1 - \beta_1) - (\mu_0 + \beta_0)}{\sqrt{\sigma^2 + \gamma^2}}\\
&s.t.\quad (1 - \tilde{p})\beta_0^2 + \tilde{p}\beta_1^2 + \gamma^2 \le D, \\
&\beta_0, \beta_1, \gamma \ge 0
    \end{flalign}
    \textbf{Proof:} For the same approximation of the Sibson mutual information we made in equation (\ref{eq:sibMIapproximation}), we can calculate the corresponding $z_0$ when
    \begin{flalign}
&\Big(\frac{\tilde{p}}{1 - \tilde{p}}\Big)^{\frac{1}{\alpha}} = \frac{\exp({-\frac{1}{2}\frac{(z - \mu_1^{'})^2}{\sigma^2 + \gamma^2}})}{\exp({-\frac{1}{2}\frac{(z - \mu_0^{'})^2}{\sigma^2 + \gamma^2}})}
    \end{flalign}
    Solving the above for $z_0$ gives
    \begin{flalign}
&z_0 = \frac{(\sigma^2 + \gamma^2)\frac{1}{\alpha}\log(\frac{\tilde{p}}{1 - \tilde{p}})}{\mu_1^{'} - \mu_0^{'}} + \frac{\mu_1^{'} + \mu_0^{'}}{2}
    \end{flalign}
    Therefore the optimization of $I_{\alpha}(C;Z)$ is monotonically increasing in $\frac{\mu_1^{'} - \mu_0^{'}}{\sqrt{\sigma^2 + \gamma^2}}$. Then
    \begin{flalign}
&(\beta_0^*, \beta_1^*, \gamma^*) = \argmin_{\beta_0, \beta_1, \gamma} \frac{\mu_1^{'} - \mu_0^{'}}{\sqrt{\sigma^2 + \gamma^2}}\\
&= \argmin_{\beta_0, \beta_1, \gamma}\frac{(\mu_1 - \beta_1) - (\mu_0 + \beta_0)}{\sqrt{\sigma^2 + \gamma^2}} \\
&= \argmin_{\beta, \gamma}\frac{(\mu_1 - \mu_0 - \beta)}{\sqrt{\sigma^2 + \gamma^2}}, \beta = \beta_0+\beta_1 \label{eq:SibMIindependent noise final opt}
    \end{flalign}
The Hessian of (\ref{eq:SibMIindependent noise final opt}) may be computed as
\begin{flalign}
&f(\beta, \gamma) = \frac{(\mu_1 - \mu_0 - \beta)}{\sqrt{\sigma^2 + \gamma^2}}, \quad \nabla^2 f = \begin{bmatrix}
\frac{\partial^2 f}{\partial \beta^2} & \frac{\partial^2 f}{\partial \beta \partial \gamma} \\
\frac{\partial^2 f}{\partial \gamma \partial \beta } & \frac{\partial^2 f}{ \partial \gamma^2}
\end{bmatrix}\\
&\frac{\partial^2 f}{\partial \beta^2} = 0\\
&\frac{\partial^2 f}{\partial \beta \partial \gamma} = \frac{\partial^2 f}{\partial \gamma \partial \beta} = \frac{\gamma}{(\sigma^2 + \gamma^2)^{\frac{3}{2}}}\\
&\frac{\partial^2 f}{ \partial \gamma^2} = (\mu_1 - \mu_0 - \beta)(-\frac{1}{2})(\sigma^2 + \gamma^2)^{-\frac{3}{2}}[(-\frac{3}{2})\frac{4 \gamma^2}{\sigma^2 + \gamma^2} + 2]
\end{flalign}
The determinant of the Hessian of (\ref{eq:SibMIindependent noise final opt}) is always non-positive, thus the optimization problem is non-convex in $\beta, \gamma$, and global optimum are not known.
\qedsymbol
\end{theorem}

\section{Data driven approach for maximal information leakage}
\label{sec: data driven approach}
\subsection{Model overview}
\par
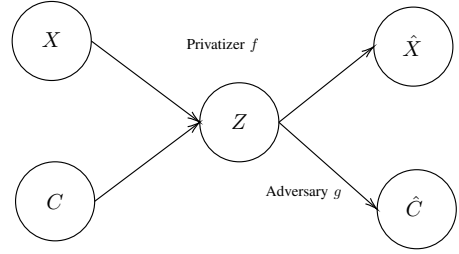
\begin {figure}
\centering
\resizebox {0.7\columnwidth} {!} {
\tikzset{every picture/.style={line width=0.75pt}} 

\tikzset{every picture/.style={line width=0.75pt}} 

\begin{tikzpicture}[x=0.75pt,y=0.75pt,yscale=-1,xscale=1]

\draw   (114,76) .. controls (114,56.67) and (129.67,41) .. (149,41) .. controls (168.33,41) and (184,56.67) .. (184,76) .. controls (184,95.33) and (168.33,111) .. (149,111) .. controls (129.67,111) and (114,95.33) .. (114,76) -- cycle ;
\draw   (117,218) .. controls (117,198.67) and (132.67,183) .. (152,183) .. controls (171.33,183) and (187,198.67) .. (187,218) .. controls (187,237.33) and (171.33,253) .. (152,253) .. controls (132.67,253) and (117,237.33) .. (117,218) -- cycle ;
\draw   (280,148) .. controls (280,128.67) and (295.67,113) .. (315,113) .. controls (334.33,113) and (350,128.67) .. (350,148) .. controls (350,167.33) and (334.33,183) .. (315,183) .. controls (295.67,183) and (280,167.33) .. (280,148) -- cycle ;
\draw   (434,81) .. controls (434,61.67) and (449.67,46) .. (469,46) .. controls (488.33,46) and (504,61.67) .. (504,81) .. controls (504,100.33) and (488.33,116) .. (469,116) .. controls (449.67,116) and (434,100.33) .. (434,81) -- cycle ;
\draw   (437,225) .. controls (437,205.67) and (452.67,190) .. (472,190) .. controls (491.33,190) and (507,205.67) .. (507,225) .. controls (507,244.33) and (491.33,260) .. (472,260) .. controls (452.67,260) and (437,244.33) .. (437,225) -- cycle ;
\draw    (184,76) -- (278.4,146.8) ;
\draw [shift={(280,148)}, rotate = 216.87] [color={rgb, 255:red, 0; green, 0; blue, 0 }  ][line width=0.75]    (10.93,-3.29) .. controls (6.95,-1.4) and (3.31,-0.3) .. (0,0) .. controls (3.31,0.3) and (6.95,1.4) .. (10.93,3.29)   ;

\draw    (187,218) -- (278.4,149.2) ;
\draw [shift={(280,148)}, rotate = 503.03] [color={rgb, 255:red, 0; green, 0; blue, 0 }  ][line width=0.75]    (10.93,-3.29) .. controls (6.95,-1.4) and (3.31,-0.3) .. (0,0) .. controls (3.31,0.3) and (6.95,1.4) .. (10.93,3.29)   ;

\draw    (350,148) -- (432.44,82.25) ;
\draw [shift={(434,81)}, rotate = 501.42] [color={rgb, 255:red, 0; green, 0; blue, 0 }  ][line width=0.75]    (10.93,-3.29) .. controls (6.95,-1.4) and (3.31,-0.3) .. (0,0) .. controls (3.31,0.3) and (6.95,1.4) .. (10.93,3.29)   ;

\draw    (350,148) -- (435.5,223.67) ;
\draw [shift={(437,225)}, rotate = 221.51] [color={rgb, 255:red, 0; green, 0; blue, 0 }  ][line width=0.75]    (10.93,-3.29) .. controls (6.95,-1.4) and (3.31,-0.3) .. (0,0) .. controls (3.31,0.3) and (6.95,1.4) .. (10.93,3.29)   ;

\draw (149,75) node [scale=1.44] [align=left] {$\displaystyle X$};
\draw (152,218) node [scale=1.44] [align=left] {$\displaystyle C$};
\draw (315,147) node [scale=1.44] [align=left] {$\displaystyle Z$};
\draw (468,80) node [scale=1.44] [align=left] {$\displaystyle \hat{X}$};
\draw (470,222) node [scale=1.44] [align=left] {$\displaystyle \hat{C}$};
\draw (372,210.5) node  [align=left] {Adversary $\displaystyle g$};
\draw (300.5,79.5) node  [align=left] {Privatizer $\displaystyle f$};

\end{tikzpicture}

}
\caption{Graph representation of the adversarial privatization model}
\label{fig: graph representation}
\end{figure}

 Given a data set consisting of N pairs of $(X, C)$ in $\{(X^{(n)}, C^{(n)})\}_{n=1}^N$, the problem is to find some (randomized) mapping $(X, C) \rightarrow Z$ such that the privatized representation $Z$ leaks as little information as possible with regards to the private variable $C$. The data $X$ is assumed to be continuous, and the private variable $C$ is a discrete variable correlated with $X$ with $G$ different possible values, often the class which $X$ belongs to. We use Sibson mutual information of order 20 in our experiments. This approximation is sufficiently close to the Sibson mutual information at order $\infty$ and do not result in numerical over/underflow during the optimization. In order to learn the mapping, we use neural networks to parameterize the adversary $g$ and privatizer $f$ in an auto-encoding model shown in Fig \ref{fig: graph representation}.
\par The presence of an adversary is to emulate an environment where the released data is gathered by an adversary, so the privatizer is encouraged to learn mappings based on a privacy metric to prevent the adversary from inferring with high accuracy. Having a trained adversary also implies that the adversary's posterior estimates $P(\hat{C}|Z)$ are close to the true posterior, allowing us to make an approximation in the calculation of empirical Sibson mutual information.
\par The adversary is trained to make inferences on the private variable, and the privatizer is trained to minimize the privacy metric and adhere to a distortion budget. For neural network privatizers, the privatizer $f(x, c) = (f_{\mu}(x,c), f_{\Sigma}(x,c))$ takes data pairs $(x,c)$ as input, and outputs the parameters of the conditional $Z$ distribution $P(Z|X,C)$. We've chosen the conditional $Z$ to be Gaussian because we believe it is a flexible distribution and allows for sampling with the method in \cite{KingmaWellingAutoencodingVB}. $S$ samples of $Z$ are generated as inputs to the adversary using the reparameterization trick from \cite{KingmaWellingAutoencodingVB}. Another approach to modeling the conditional latent distribution released by the privatizer is demonstrated in \cite{MakhzaniAdvAutoencoders}. The privatizer also reconstructs $\hat{X}$ from the samples of $Z$ to let us compute the reconstruction error component in its loss function. The adversary $g(z)$ outputs predictions for $C$ in the vector $P(\hat{C}|Z)$ given the average over the $S$ samples of $Z$.
\par For synthetic data, we also conduct experiments with (noisy) affine encoders, but the adversary is still represented by a neural network. When we are using neural networks to parameterize the encoder in experiments, we measure distortion as the average reconstruction error by default
\begin{flalign}
&\mathop{{}\mathbb{E}}_x[d(X, \hat{X})] = \frac{1}{N}\displaystyle\sum_n^N d(x^{(n)}, \hat{x}^{(n)}) \le D\\
&d(x,\hat{x}) = ||x - \hat{x}||_2^2 \label{eq: L2distortion}
\end{flalign}
When using (noisy) affine transformations in the encoder, we measure the distortion as 
\begin{flalign} \label{eq: affine distortion from magnitude}
    (1-\tilde{p})\beta_0^2 + \tilde{p}\beta_1^2 \le D
\end{flalign}
where the parameters of the encoder are $\beta_0, \beta_1$ and the transform is from equation (\ref{eq: affine transform}), or 
\begin{flalign} \label{eq: noisy affine distortion from magnitude}
    (1 - \tilde{p})\beta_0^2 + \tilde{p}\beta_1^2 + \gamma^2 \le D
\end{flalign}
where the parameters of the encoder are $\beta_0, \beta_1, \gamma$ and the transform is from equation (\ref{eq:affine with noise}).
\par For experiments with the synthetic Gaussian data and (noisy) affine transformations, measuring the expected L2 distance from the reconstruction $\hat{X} = Z$ to the original $X$ is equivalent to measuring $(1-\tilde{p})\beta_0^2 + \tilde{p}\beta_1^2$ up to a scaling factor in $D$. This is due to the fact that on average, the expected distortion for affine transformations from equation (\ref{eq: L2distortion}) is
\begin{flalign} \label{eq: L2 distortion is equal to square of distortion params for affine}
&\frac{1}{N}\displaystyle\sum_n^N d(x^{(n)}, \hat{x}^{(n)}) = \frac{1}{N} \displaystyle\sum_n^N ||x^{(n)} - \hat{x}^{(n)}||_2^2\\
&= \frac{1}{N} \displaystyle\sum_n^N ||x^{(n)} - z^{(n)}||_2^2 = (1 - \tilde{p})\beta_0^2 + \tilde{p}\beta_1^2
\end{flalign}
For noisy affine transformations, it is 
\begin{flalign}\label{eq: L2 distortion is equal to square of distortion params for affine noisy}
&\frac{1}{N} \displaystyle\sum_n^N ||x^{(n)} - \hat{x}^{(n)}||_2^2 = \frac{1}{N} \displaystyle\sum_n^N ||x^{(n)} - z^{(n)}||_2^2 \\
&= (1 - \tilde{p})\beta_0^2 + \tilde{p}\beta_1^2 + \gamma^2
\end{flalign}
Thus we use the distortion metric in equations (\ref{eq: affine distortion from magnitude})(\ref{eq: noisy affine distortion from magnitude}). However, for neural networks learning non-linear mappings of the private representation, it is more appropriate to measure the distortion in terms of the average reconstruction error from equation (\ref{eq: L2distortion}).

In the data set we have $N$ pairs of data, and throughout training we assume the adversary is trained and generates near-optimal posterior probability vectors to classify the private label. The posterior probability is used as part of the empirical loss function discussed below. Both the adversary $f$ and the privatizer $g$ are neural networks each parameterized by \begin{math}
\theta_p
\end{math} and \begin{math}
\theta_a
\end{math}.

The goal in the data-driven approach is to learn a mapping for data pairs such that the Sibson mutual information is low, subject to a distortion constraint. The optimal parameters for $\theta_p, \theta_a$ are found through an iterative alternating training algorithm to keep the adversary optimal for each iteration of optimization for the privatizer over the empirical approximation of the Sibson mutual information, which are further discussed.

\subsection{Empirical loss}
In this section we will discuss our approximation of the maximal information leakage as a metric for a data-driven privatization model, and compare it with mutual information. Calculating the Sibson mutual information requires knowledge of the posterior distribution $P(C|Z)$ which is not easily accessible, but due to the presence of a trained adversary, we have access to the MAP adversary's posterior estimate of $\hat{C}$ after the observation of $Z$. Along with a predetermined prior probability of $C$ which the MAP adversary also has access to, we may approximate the Sibson mutual information by using the empirical estimate of the posterior on $\hat{C}$.
\small
\begingroup
\allowdisplaybreaks
\begin{flalign}
&I_{\alpha}(C;Z) = \frac{\alpha}{\alpha-1}\log(\displaystyle\int_z (\displaystyle\sum_c^G P_{Z|C}^{\alpha}(z|c)P_C(c))^{1/\alpha} dz) \\
&= \frac{\alpha}{\alpha-1}\log(\displaystyle\int_z (\displaystyle\sum_c^G P_{C|Z}^{\alpha}(c|z)P_Z^{\alpha}(z)P_C(c)/P_C^{\alpha}(c))^{1/\alpha} dz) \\
&= \frac{\alpha}{\alpha-1}\log(\displaystyle\int_z (\displaystyle\sum_c^G P_{C|Z}^{\alpha}(c|z) P_C^{1 - \alpha}(c))^{1/\alpha} P_Z(z) dz) \\
&= \frac{\alpha}{\alpha-1}\log\Big(\displaystyle\int_z \big(\displaystyle\sum_c^G P_{C|Z}^{\alpha}(c|z) P_C^{1 - \alpha}(c)\big)^{1/\alpha} \\
&\quad * \displaystyle\int_x P_{Z|X}(z|x) P_X(x) dxdz\Big) \nonumber\\
&\approx \frac{\alpha}{\alpha-1}\log(\displaystyle\sum_n^N \frac{1}{N} \displaystyle\int_z P_{Z|X}(z|x_n)\\
&\quad (\displaystyle\sum_c^G P_{C|Z}^{\alpha}(c_n|z)  P_C^{1 - \alpha}(c_n))^{1/\alpha} dz)  \nonumber\\
&\approx \frac{\alpha}{\alpha-1}\log(\displaystyle\sum_n^N \frac{1}{N}(\displaystyle\sum_i^S \frac{1}{S}(\displaystyle\sum_c^G P_{C|Z}^{\alpha}(c_n|z_{i,n})P_C^{1-\alpha}(c_n))^{1/\alpha})) \\
&\approx \frac{\alpha}{\alpha-1}\log(\displaystyle\sum_n^N \frac{1}{N}(\displaystyle\sum_i^S \frac{1}{S}(\displaystyle\sum_c^G\\
&\quad P_{\hat{C}|Z}^{\alpha}(\hat{c}_n|z_{i,n}, \theta_a, \theta_p)P_C^{1-\alpha}(c_n))^{1/\alpha})) \nonumber\\
&= \frac{\alpha}{\alpha-1}\log(\displaystyle\sum_n^N \frac{1}{N}(\displaystyle\sum_i^S \frac{1}{S}(\displaystyle\sum_c^G\\
& \quad (\frac{P_{\hat{C}|Z}(\hat{c}_n|z_{i,n}, \theta_a, \theta_p)}{P_C(c_n)})^{\alpha}P_C(c_n))^{1/\alpha})) \nonumber\\
&= \frac{\alpha}{\alpha-1}\log(\displaystyle\sum_n^N \frac{1}{N}(\displaystyle\sum_i^S \frac{1}{S}\mathbb{E}_{C}[ (\frac{P_{\hat{C}|Z}(\hat{c}_n|z_{i,n}, \theta_a, \theta_p)}{P_C(c_n)})^{\alpha}]^{1/\alpha}))\\
&\doteq j(\hat{C}, Z, \theta_a, \theta_p)
\label{eq: sibMI empirical}
\end{flalign}
\endgroup
\normalsize
Starting with the definition of the Sibson mutual information for a continuous $Z$ and discrete $C$ summed over its $G$ classes, we use Bayes' rule and then expand $P(Z)$ into $\sum_x P(Z|X)P(X)$. The summation over $P(X)$ is approximated by the sum over the $N$ data points. Instead of integration over the support of $Z$, we average over samples of the conditional $Z$ distribution due to the fact that outputting and summing over the classes $C$ is only available through the adversary, and the adversary takes discrete points of $Z$ as input. We use the estimated posterior on $C$ in the approximation because for every iteration of optimization over the privatizer, the adversary is trained and can produce an estimate $P(\hat{C}|Z)$ that is close to the true distribution.
In comparison, with mutual information, we have:
\begingroup
\allowdisplaybreaks
\begin{flalign}
&I(C;Z) = \displaystyle\int_z \displaystyle\sum_c^G P_{Z, C}(z, c) \log\Big(\frac{P_{C|Z}(c|z)}{P_C(c)}\Big)dz\\
&= \displaystyle\sum_n^N \frac{1}{N} \displaystyle\int_z \displaystyle\sum_c^G P_{C|Z}(c_n|z) P_{Z|X}(z|x_n) \log\Big(\frac{P_{C|Z}(c_n|z)}{P_C(c_n)}\Big)dz\\
&\approx \displaystyle\sum_n^N \frac{1}{N}\displaystyle\sum_i^S \frac{1}{S} \displaystyle\sum_c^G P_{\hat{C}|Z}(\hat{c}_n|z_{i,n}, \theta_a, \theta_p)\\ 
&\log\Big(\frac{P_{\hat{C}|Z}(\hat{c}_n|z_{i,n}, \theta_a, \theta_p)}{P_C(c_n)} \Big)
\label{eq: empirical KL}
\end{flalign}
\endgroup
We can interpret the mutual information as the Kullback-Leibler divergence between the posterior estimate of $C$ given $Z$ and the prior estimate of $C$, and use it as a comparative metric denoted as "MI" in experiments with MNIST and FERG data. The Sibson mutual information estimate allows us to design an adversarial model to minimize it as an objective function when learning the privacy mapping. Our model consists of encoder and decoders parameterized by neural networks (represented in Fig \ref{fig: graph representation}), and the training procedure are discussed in the following section.

\subsection{Alternating training algorithm}
The encoder acts as a privatizer operating under the assumption that an optimal adversary is available, and optimizes to minimize the Sibson mutual information \begin{math}
I_{\alpha}(C;Z)
\end{math} subject to the distortion budget $D$. They are each parameterized by \begin{math}
\theta_p
\end{math} and \begin{math}
\theta_a
\end{math} respectively. The models are trained for 2000 epochs for synthetic data, 200 epochs for MNIST data set, and 200 for FERG data using the Adam optimizer\cite{adam}. Each epoch consists of a pass over all data points in the training set divided in mini-batches of size $M$ for a total of $N/M$ iterations, and the empirical Sibson mutual information/cross-entropy is computed below for each mini-batch. The adversary is trained with its objective equation (\ref{eq: minibatch L_a}) for $k = 20$ iterations for each iteration of training for the privatizer. The objective functions for each component for each mini-batch of size $M$ at iteration $t$ are:
\begingroup
\allowdisplaybreaks
\begin{flalign}
&\label{eq: minibatch L_a} L_a(\theta_p^t, \theta_a^t) = \\
&\frac{1}{M} \displaystyle \sum_{n=1}^{M}\sum_c^G \mathbbm{1}(C^{(n)}=c)[-\log(P(\hat{C}^{(n)}=c|Z^{(n)};\theta_a^t|\theta_p^t))] \nonumber\\
&L_p(\theta_p^t, \theta_a^t, \rho_t) = j(\hat{C}^t, Z^t, \theta_a^t, \theta_p^t) \label{eq: minibatch L_p}\\
&+ \rho_t \max\Big\{0, \frac{1}{M} \displaystyle\sum_{n=1}^{M}d(\hat{x}_n, x_n) - D\Big\} \nonumber\\
&j(\hat{C}^t, Z^t, \theta_a^t, \theta_p^t) =\frac{\alpha}{\alpha-1} \cdot\\
&\log\Big(\displaystyle\sum_n^M \frac{1}{M}\big(\displaystyle\sum_i^S \frac{1}{S}(\displaystyle\sum_c^G (\frac{P_{\hat{C}|Z}(\hat{c}_n|z_{i,n}; \theta_p^t|\theta_a^t)}{P_C(c_n)})^{\alpha}P_C(c_n))^{1/\alpha}\big)\Big) \nonumber
\end{flalign}
\endgroup
The adversary's loss is dependent on the posterior estimate $P(\hat{C}|Z;\theta_a^t|\theta_p^t)$, conditioned on the privatizer network's generated representation of $Z$; the privatizer's loss depends on the posterior estimate $P(\hat{C}|Z;\theta_p^t|\theta_a^t)$, conditioned on the adversary network's prediction of $\hat{C}$. The empirical estimate of the Sibson mutual information is derived from equation (\ref{eq: sibMI empirical}) for a mini-batch of size $M$.

For synthetic data, the distortion measure is the distortion budget 
\begin{flalign}
\mathop{{}\mathbb{E}}_x[d(X, \hat{X})] = (1-\tilde{p})\beta_0^2 + \tilde{p}\beta_1^2
\end{flalign} for an affine privatizer, and 
\begin{flalign}
\mathop{{}\mathbb{E}}_x[d(X, \hat{X})] = (1 - \tilde{p})\beta_0^2 + \tilde{p}\beta_1^2 + \gamma^2
\end{flalign}
for a noisy affine privatizer. For synthetic data using a neural network privatizer and for real-world data, we use the L2 distance specified in equation (\ref{eq: L2distortion}), and the penalty coefficient \begin{math}
\rho_t
\end{math} increases with the number of iterations \begin{math}
t
\end{math}.
The algorithm is given in Algorithm \ref{alg: alternating training algorithm}.
\begin{algorithm}
    \caption{Alternate training for privacy-preserving adversarial model}
    \label{alternatealgo}
    \begin{algorithmic}
    \REQUIRE $M, S, N, k, D, \{c_i\}, \{x_i\}$
    \ENSURE $\theta_p^{T}, \theta_a^{T}$
    \STATE $\theta_p^0 \leftarrow \mathcal{N}(0, I), \theta_a^0 \leftarrow \mathcal{N}(0, I), t = 0$
    \WHILE{$t \le T$}
    \STATE $\rho_t = \frac{10t}{T} + 1$
    \STATE $\theta_a^{t,0} \leftarrow \theta_a^t$
    \FOR{$(j = 0; j < k; j++)$}
    \STATE $\theta_a^{t,j+1} \leftarrow f_{Adam}(\nabla_{\theta_a}L_a(\theta_p^t, \theta_a^{t,j}, \rho_t))$
    \COMMENT{$f_{Adam}$ is the output from one update of the Adam optimizer on the adversary's loss component}
    \ENDFOR
    \STATE $\theta_a^{t+1} \leftarrow \theta_a^{t,k-1}$
    \STATE $\theta_p^{t+1} \leftarrow g_{Adam}(\nabla_{\theta_p}L_p(\theta_p^t, \theta_a^{t+1}, \rho_t))$
    \COMMENT{$g_{Adam}$ is the output from one update of the Adam optimizer on the privatizer's loss component}
    \STATE $t \leftarrow t + 1$
    \ENDWHILE
    \RETURN $\theta_p^{T}, \theta_a^{T}$
    \end{algorithmic}
    \label{alg: alternating training algorithm}
\end{algorithm}

The reason for iterating over the training of the adversary $k$ times for each iteration of the privatizer training is to ensure that the adversary is sufficiently trained, and produces the posterior probabilities of the private labels that are able to classify them well. The inner loop updates the parameters of the adversary network $k$ times according to the adversary loss $L_a$ to maintain a trained adversary for every time the privatizer's parameters are updated. Therefore the privatizer can operate under the assumption that a trained adversary is present, as specified in the design of this model. In practice we used $k=20$ for training for synthetic data and $k=10$ for MNIST and FERG data. The following section demonstrates the use of Sibson mutual information as the privacy metric in an adversarial model with synthetic and real-world data and our experimental results.

\section{Experiments} \label{sec: experiments section}
We conduct experiments with 1-D synthetic data drawn from a Bernoulli-Gaussian distribution, the MNIST data set and the FERG data set, where the private variable is the class of the data point, and this section reports the results which show that Sibson mutual information offers equivalent or favorable performance in comparison with mutual information. All of the models below are trained with stochastic gradient descent of the loss function with mini-batches of data using the Adam optimizer\cite{adam} on default hyper-parameter settings with Algorithm \ref{alg: alternating training algorithm}. 

\subsection{Synthetic data}
Synthetic data is generated by drawing from a Bernoulli prior distribution with $\tilde{p} = 1 - \tilde{p} = 0.5$ for the class of each data point, and conditioned on the class for each point, the $X$ variable is drawn from a Gaussian distribution with parameters $\mathcal{N}(3, 1)$ and $\mathcal{N}(-3, 1)$ for 15000 points. Of those points, 10000 are used for training, and 5000 are used for validation. For the synthetic data, we consider the encoder as affine (section \ref{sec:affine sibmi sec}), affine with noise (section \ref{sec:Affine noisy extension}), or a fully connected neural network with layers of (4, 2) hidden units which map a 1 dimensional X into the two parameters of the 1 dimensional Z which is distributed as a Gaussian. We average over a sample of 12 points from the Z distribution and feed to the decoder, a fully connected neural network with layers of (4,2) hidden units in the reconstruction and inference branch respectively. The outputs of the decoder are $\hat{X}$ and $P(C|Z)$, where the adversary aims to minimize the cross-entropy loss, the encoder aims to minimize the privacy metric, subject to a reconstruction constraint. For optimization, we use the Adam optimizer\cite{adam} on our algorithm with a learning rate of \begin{math}
10^{-3}
\end{math} and a mini-batch size of \begin{math}
M = 500
\end{math} over $1000$ epochs.

\par
\begin{figure}[thpb]
      \centering
      \includegraphics[width=0.4 \textwidth]{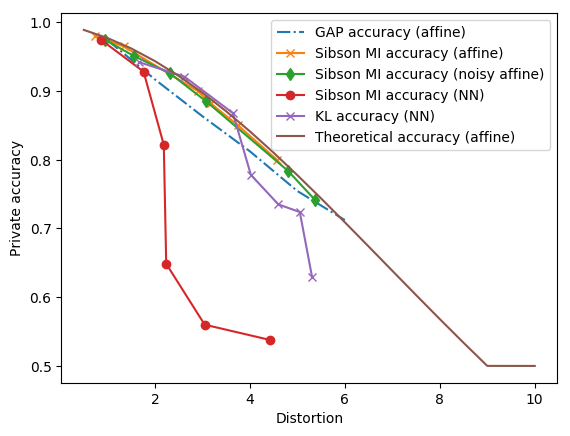}
      \caption{Synthetic Gaussian data adversary accuracy rate vs distortion budget}
      \label{fig:synth}
\end{figure}

 We implement a wide range of encoders using both Sibson mutual information and mutual information as the privacy metric for the synthetic data set described previously. With affine transformations, we show the theoretical MAP adversary accuracy based on the solutions of optimizing for maximal information leakage (equation (\ref{eq: MAP accuracy expression})(\ref{eq:affine solution 1})(\ref{eq:affine solution 2})), which we demonstrated was equal to the solutions with MAP adversary accuracy and Sibson mutual information (Corollary \ref{cor: corollary connecting the solutions for affine transformation}). This is the theoretical baseline that data-driven approaches aim to approximate. We then implement a data-driven model with an affine encoder and a neural network adversary with two layers of $(4, 2)$ hidden units with ReLU activations and show the MAP adversary's accuracy over various distortion budgets as measured by equation (\ref{eq: affine distortion from magnitude}). This model follows Algorithm \ref{alg: alternating training algorithm} and uses equation (\ref{eq: minibatch L_a})(\ref{eq: minibatch L_p}) as the losses. As a comparison we include the MAP adversary accuracy for the data-driven GAP framework \cite{CGAPpaper} which minimizes the mutual information for the same affine transformation. We also implement a noisy affine encoder described by the transform (\ref{eq:affine with noise}) that optimizes the Sibson mutual information, and plot the adversary accuracy over the distortion budget. This model variant is trained with Algorithm \ref{alg: alternating training algorithm}, the adversary uses the loss in equation (\ref{eq: minibatch L_a}), and the privatizer uses equation (\ref{eq: minibatch L_p}) but with equation (\ref{eq: L2 distortion is equal to square of distortion params for affine noisy}) in its distortion budget term. From Fig \ref{fig:synth} we can see that the data-driven models with Sibson mutual information achieve adversary accuracies that closely approximate the theoretical accuracy, and the discrepancy between the GAP framework's accuracy and the theoretical one may be due to their approach not training the adversary sufficiently.

\par Finally, we implement a simple neural network encoder and adversary with Sibson mutual information as the privacy metric. The encoder takes in $(X,C)$ pairs as input and has two layers of $(4, 2)$ hidden units with ReLU activations, while the decoder has two branches of $(4, 2)$ hidden units with ReLU activations in each branch, and outputs $(\hat{X}, P(\hat{C}|Z))$. This is again trained with Algorithm \ref{alg: alternating training algorithm} and uses the losses in equation (\ref{eq: minibatch L_a})(\ref{eq: minibatch L_p}). We compare this with the mutual information metric using the same model, but the empirical mutual information from equation (\ref{eq: empirical KL}) instead of (\ref{eq: sibMI empirical}) when calculating the privatizer loss from equation (\ref{eq: minibatch L_p}). The adversary's accuracy for both metrics are plotted in Fig \ref{fig:synth} labeled as "(NN)". We see that Sibson mutual information offers greater privacy than mutual information for the same distortion, as measured by the lower adversary accuracy and both are lower than the (noisy) affine transformations due to the fact that the mapping learned by a neural network encoder is more complex.

\par The actual values of the distortion and adversary accuracy can be seen in Table \ref{tab:synth_table}. Due to the non-linearity of our data-driven model with a neural network encoder, we conduct further experiments to illustrate the viability of using maximal information leakage as the privacy metric.

\begin{table}[h]
  \caption{Synthetic data results compared with the GAP framework\cite{CGAPpaper}, distortion vs. adversary accuracy}
  \label{tab:synth_table}
\resizebox{1.05\columnwidth}{!}{
\begin{tabular}{|l|l|l|l|l|l|ll}
\cline{1-6}
\begin{tabular}[c]{@{}l@{}}Distortion \\ Budget\end{tabular} & \begin{tabular}[c]{@{}l@{}}GAP \\ accuracy\end{tabular} & \begin{tabular}[c]{@{}l@{}}Experimental\\ Distortion\end{tabular} & \begin{tabular}[c]{@{}l@{}}Sibson MI\\ accuracy\\ (Affine)\end{tabular} & \begin{tabular}[c]{@{}l@{}}Experimental\\ Distortion\end{tabular} & \begin{tabular}[c]{@{}l@{}}Sibson MI \\ accuracy\\ (Affine \\ with noise)\end{tabular} &  &  \\ \cline{1-6}
1                                                            & 0.9742                                                  & 0.738                                                             & 0.980                                                                   & 0.936                                                             & 0.975                                                                                  &  &  \\ \cline{1-6}
2                                                            & 0.9169                                                  & 1.340                                                             & 0.965                                                                   & 1.56                                                              & 0.951                                                                                  &  &  \\ \cline{1-6}
3                                                            & 0.8633                                                  & 2.904                                                             & 0.900                                                                   & 2.31                                                              & 0.926                                                                                  &  &  \\ \cline{1-6}
4                                                            & 0.8123                                                  & 3.174                                                             & 0.882                                                                   & 3.08                                                              & 0.885                                                                                  &  &  \\ \cline{1-6}
5                                                            & 0.7545                                                  & 3.750                                                             & 0.850                                                                   & 4.80                                                              & 0.784                                                                                  &  &  \\ \cline{1-6}
6                                                            & 0.7122                                                  & 4.570                                                             & 0.800                                                                   & 5.38                                                              & 0.741                                                                                  &  &  \\ \cline{1-6}
                                                             &                                                         & \begin{tabular}[c]{@{}l@{}}Experimental\\ Distortion\end{tabular} & \begin{tabular}[c]{@{}l@{}}Sibson MI \\ Accuracy (NN)\end{tabular}      & \begin{tabular}[c]{@{}l@{}}Experimental \\ Distortion\end{tabular} & \begin{tabular}[c]{@{}l@{}}MI accuracy\\ (NN)\end{tabular}                             &  &  \\ \cline{1-6}
                                                             &                                                         & 0.867                                                             & 0.9745                                                                  & 1.67                                                              & 0.942                                                                                  &  &  \\ \cline{1-6}
                                                             &                                                         & 1.76                                                              & 0.9283                                                                  & 2.62                                                              & 0.921                                                                                  &  &  \\ \cline{1-6}
                                                             &                                                         & 2.19                                                              & 0.8218                                                                  & 3.64                                                              & 0.868                                                                                  &  &  \\ \cline{1-6}
                                                             &                                                         & 2.24                                                              & 0.6486                                                                  & 4.02                                                              & 0.778                                                                                  &  &  \\ \cline{1-6}
                                                             &                                                         & 3.05                                                              & 0.5600                                                                  & 4.60                                                              & 0.735                                                                                  &  &  \\ \cline{1-6}
                                                             &                                                         & 4.43                                                              & 0.5377                                                                  & 5.05                                                              & 0.724                                                                                  &  &  \\ \cline{1-6}
                                                             &                                                         &                                                                   &                                                                         & 5.31                                                              & 0.629                                                                                  &  &  \\ \cline{1-6}
\end{tabular}
}
\end{table}

\subsection{MNIST data}
\begin{figure}        
\resizebox {\columnwidth} {!} {

\tikzset{every picture/.style={line width=0.75pt}} 
\begin{tikzpicture}[x=0.75pt,y=0.75pt,yscale=-1,xscale=1]

\draw   (42,84) -- (74,52) -- (90,52) -- (90,218) -- (58,250) -- (42,250) -- cycle ; \draw   (90,52) -- (58,84) -- (42,84) ; \draw   (58,84) -- (58,250) ;
\draw   (105,93.98) -- (123.98,75) -- (140,75) -- (140,191.02) -- (121.02,210) -- (105,210) -- cycle ; \draw   (140,75) -- (121.02,93.98) -- (105,93.98) ; \draw   (121.02,93.98) -- (121.02,210) ;
\draw   (150,102) -- (172,80) -- (187,80) -- (187,158) -- (165,180) -- (150,180) -- cycle ; \draw   (187,80) -- (165,102) -- (150,102) ; \draw   (165,102) -- (165,180) ;
\draw   (196,106.72) -- (211.72,91) -- (220,91) -- (220,144.28) -- (204.28,160) -- (196,160) -- cycle ; \draw   (220,91) -- (204.28,106.72) -- (196,106.72) ; \draw   (204.28,106.72) -- (204.28,160) ;
\draw   (242.63,70) -- (250,70) -- (250,173.47) -- (242.63,173.47) -- cycle ;
\draw   (272.63,70) -- (280,70) -- (280,173.47) -- (272.63,173.47) -- cycle ;
\draw   (300,70) -- (307.37,70) -- (307.37,173.47) -- (300,173.47) -- cycle ;
\draw   (406,65.72) -- (421.72,50) -- (430,50) -- (430,103.28) -- (414.28,119) -- (406,119) -- cycle ; \draw   (430,50) -- (414.28,65.72) -- (406,65.72) ; \draw   (414.28,65.72) -- (414.28,119) ;
\draw   (440,52) -- (462,30) -- (477,30) -- (477,108) -- (455,130) -- (440,130) -- cycle ; \draw   (477,30) -- (455,52) -- (440,52) ; \draw   (455,52) -- (455,130) ;
\draw   (485,33.98) -- (503.98,15) -- (520,15) -- (520,131.02) -- (501.02,150) -- (485,150) -- cycle ; \draw   (520,15) -- (501.02,33.98) -- (485,33.98) ; \draw   (501.02,33.98) -- (501.02,150) ;
\draw   (530,34) -- (562,2) -- (578,2) -- (578,168) -- (546,200) -- (530,200) -- cycle ; \draw   (578,2) -- (546,34) -- (530,34) ; \draw   (546,34) -- (546,200) ;
\draw   (386,82) -- (396,82) -- (396,150) -- (386,150) -- cycle ;
\draw    (250,120) -- (268,120) ;
\draw [shift={(270,120)}, rotate = 180] [color={rgb, 255:red, 0; green, 0; blue, 0 }  ][line width=0.75]    (10.93,-3.29) .. controls (6.95,-1.4) and (3.31,-0.3) .. (0,0) .. controls (3.31,0.3) and (6.95,1.4) .. (10.93,3.29)   ;

\draw    (280,120) -- (298,120) ;
\draw [shift={(300,120)}, rotate = 180] [color={rgb, 255:red, 0; green, 0; blue, 0 }  ][line width=0.75]    (10.93,-3.29) .. controls (6.95,-1.4) and (3.31,-0.3) .. (0,0) .. controls (3.31,0.3) and (6.95,1.4) .. (10.93,3.29)   ;

\draw    (80,150) -- (98,150) ;
\draw [shift={(100,150)}, rotate = 180] [color={rgb, 255:red, 0; green, 0; blue, 0 }  ][line width=0.75]    (10.93,-3.29) .. controls (6.95,-1.4) and (3.31,-0.3) .. (0,0) .. controls (3.31,0.3) and (6.95,1.4) .. (10.93,3.29)   ;

\draw    (130,140) -- (148,140) ;
\draw [shift={(150,140)}, rotate = 180] [color={rgb, 255:red, 0; green, 0; blue, 0 }  ][line width=0.75]    (10.93,-3.29) .. controls (6.95,-1.4) and (3.31,-0.3) .. (0,0) .. controls (3.31,0.3) and (6.95,1.4) .. (10.93,3.29)   ;

\draw    (179,129) -- (194.37,128.53) ;
\draw [shift={(196.37,128.47)}, rotate = 538.24] [color={rgb, 255:red, 0; green, 0; blue, 0 }  ][line width=0.75]    (10.93,-3.29) .. controls (6.95,-1.4) and (3.31,-0.3) .. (0,0) .. controls (3.31,0.3) and (6.95,1.4) .. (10.93,3.29)   ;

\draw    (220,120) -- (238,120) ;
\draw [shift={(240,120)}, rotate = 180] [color={rgb, 255:red, 0; green, 0; blue, 0 }  ][line width=0.75]    (10.93,-3.29) .. controls (6.95,-1.4) and (3.31,-0.3) .. (0,0) .. controls (3.31,0.3) and (6.95,1.4) .. (10.93,3.29)   ;

\draw    (420,90) -- (438,90) ;
\draw [shift={(440,90)}, rotate = 180] [color={rgb, 255:red, 0; green, 0; blue, 0 }  ][line width=0.75]    (10.93,-3.29) .. controls (6.95,-1.4) and (3.31,-0.3) .. (0,0) .. controls (3.31,0.3) and (6.95,1.4) .. (10.93,3.29)   ;

\draw    (460,90) -- (478,90) ;
\draw [shift={(480,90)}, rotate = 180] [color={rgb, 255:red, 0; green, 0; blue, 0 }  ][line width=0.75]    (10.93,-3.29) .. controls (6.95,-1.4) and (3.31,-0.3) .. (0,0) .. controls (3.31,0.3) and (6.95,1.4) .. (10.93,3.29)   ;

\draw    (510,90) -- (528,90) ;
\draw [shift={(530,90)}, rotate = 180] [color={rgb, 255:red, 0; green, 0; blue, 0 }  ][line width=0.75]    (10.93,-3.29) .. controls (6.95,-1.4) and (3.31,-0.3) .. (0,0) .. controls (3.31,0.3) and (6.95,1.4) .. (10.93,3.29)   ;

\draw   (414,176.53) -- (421.37,176.53) -- (421.37,280) -- (414,280) -- cycle ;
\draw   (448,176.53) -- (455.37,176.53) -- (455.37,280) -- (448,280) -- cycle ;
\draw   (480,196.53) -- (490,196.53) -- (490,260) -- (480,260) -- cycle ;
\draw   (212,170.53) -- (222,170.53) -- (222,234) -- (212,234) -- cycle ;
\draw    (222.37,203.47) -- (241.72,146.36) ;
\draw [shift={(242.37,144.47)}, rotate = 468.73] [color={rgb, 255:red, 0; green, 0; blue, 0 }  ][line width=0.75]    (10.93,-3.29) .. controls (6.95,-1.4) and (3.31,-0.3) .. (0,0) .. controls (3.31,0.3) and (6.95,1.4) .. (10.93,3.29)   ;

\draw    (423.68,228.27) -- (441.68,228.27) ;
\draw [shift={(443.68,228.27)}, rotate = 180] [color={rgb, 255:red, 0; green, 0; blue, 0 }  ][line width=0.75]    (10.93,-3.29) .. controls (6.95,-1.4) and (3.31,-0.3) .. (0,0) .. controls (3.31,0.3) and (6.95,1.4) .. (10.93,3.29)   ;

\draw    (457.68,227.27) -- (475.68,227.27) ;
\draw [shift={(477.68,227.27)}, rotate = 180] [color={rgb, 255:red, 0; green, 0; blue, 0 }  ][line width=0.75]    (10.93,-3.29) .. controls (6.95,-1.4) and (3.31,-0.3) .. (0,0) .. controls (3.31,0.3) and (6.95,1.4) .. (10.93,3.29)   ;

\draw    (396.37,129.47) -- (411.37,225.29) ;
\draw [shift={(411.68,227.27)}, rotate = 261.1] [color={rgb, 255:red, 0; green, 0; blue, 0 }  ][line width=0.75]    (10.93,-3.29) .. controls (6.95,-1.4) and (3.31,-0.3) .. (0,0) .. controls (3.31,0.3) and (6.95,1.4) .. (10.93,3.29)   ;

\draw    (396.37,110.47) -- (405.15,99.05) ;
\draw [shift={(406.37,97.47)}, rotate = 487.57] [color={rgb, 255:red, 0; green, 0; blue, 0 }  ][line width=0.75]    (10.93,-3.29) .. controls (6.95,-1.4) and (3.31,-0.3) .. (0,0) .. controls (3.31,0.3) and (6.95,1.4) .. (10.93,3.29)   ;

\draw (350,120) node [scale=0.9] [align=left] {$\displaystyle  \begin{array}{{>{\displaystyle}l}}
sample\sim \\
\mathcal{N}( \mu ,\ \Sigma )
\end{array}$};
\draw (304,188) node  [align=left] {$\displaystyle \mu ,\ \Sigma $};
\draw (499,226) node [scale=0.9] [align=left] {$\displaystyle \hat{C}$};
\draw (587,99) node [scale=0.9] [align=left] {$\displaystyle \hat{X}$};
\draw (33,154) node [scale=0.9] [align=left] {$\displaystyle X$};
\draw (202,203) node [scale=0.9] [align=left] {$\displaystyle C$};
\draw (392,160) node [scale=0.9] [align=left] {$\displaystyle Z$};
\draw (50,262) node [scale=0.7] [align=left] {(28,28)};
\draw (547,210) node [scale=0.7] [align=left] {(28,28)};
\draw (216,244) node [scale=0.7] [align=left] {(10,1)};
\draw (484,272) node [scale=0.7] [align=left] {(10,1)};
\draw (389,177) node [scale=0.7] [align=left] {(128,1)};
\draw (244,61) node [scale=0.7] [align=left] {(1024)};
\draw (275,61) node [scale=0.7] [align=left] {(512)};
\draw (307,61) node [scale=0.7] [align=left] {(256)};
\draw (415,290) node [scale=0.7] [align=left] {(512)};
\draw (451,290) node [scale=0.7] [align=left] {(256)};

\end{tikzpicture}

}
\caption{MNIST data-driven privatization model, numbers in braces represent tensor dimensions}
\label{fig:mnist_model_structure}
\end{figure}
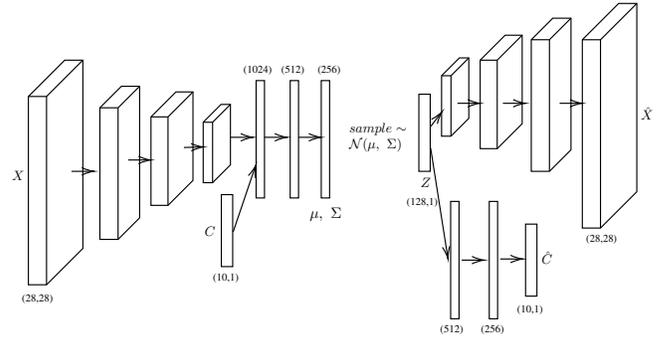

The MNIST data consists of 60000 gray-scale images of pen-written digits and their corresponding digit label, where the images are $28\times28$ binary arrays, and the digit label is a one-hot vector of length 10. 50000 data points are used for training, and 10000 are used for validation. For the MNIST data, we implemented a 3-layer convolutional neural network for the encoder, a 4-layer deconvolutional network for the decoder's reconstruction branch, and a fully connected network with two layers of (512, 256) hidden units with ReLU activation for the decoder's inference branch, as can be seen in Fig \ref{fig:mnist_model_structure}. The convolutional layers in the encoder consist of (32, 64, 128) filters of length 5, one dropout layer and two fully connected layers to output the dimensions for the parameters of $Z$. The privatized representation $Z$ is a 128 dimensional isotropic Gaussian whose parameters $(\mu_z, \Sigma_z)$ are generated by the encoder.

The deconvolutional network branch for the decoder consists of (128, 64, 32, 1) deconvolutional filters of length (3, 5, 5, 5), and the inference branch of the decoder has fully connected layers of (512, 256) hidden units with ReLU activations. The inference metric for the adversary is cross-entropy as in equation (\ref{eq: minibatch L_a}) and the privacy metrics are Sibson mutual information (equation (\ref{eq: sibMI empirical})) and mutual information (equation (\ref{eq: empirical KL})) as comparison. For optimization, we use the Adam optimizer\cite{adam} on our algorithm with a learning rate of \begin{math}
10^{-3}
\end{math} and a mini-batch size of \begin{math}
M = 500
\end{math} over $200$ epochs and $k = 20$.
\par
\begin{figure}[thpb]
      \centering
      \includegraphics[width=0.5 \textwidth]{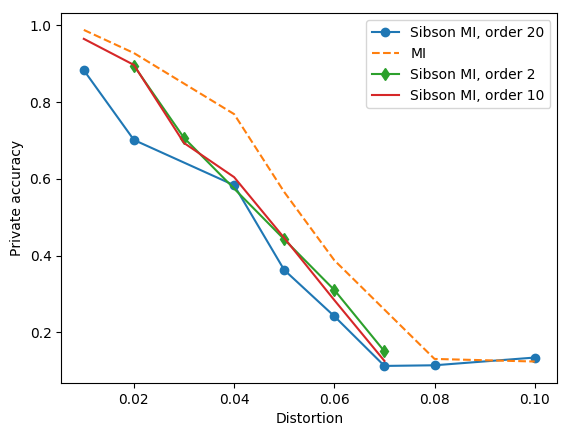}
      \caption{MNIST data adversary accuracy rate vs distortion budget of L2 reconstruction (lower is better)}
      \label{fig:mnist_L2}
\end{figure}

 As the distortion budget is increased, we can achieve various points along the privacy-utility trade-off curve, as measured by the adversary accuracy against distortion seen in Fig \ref{fig:mnist_L2}.  With a distortion budget that was enforced by an increasing penalty coefficient in equation (\ref{eq: minibatch L_p}), we were able to obtain adversary performances varying between random guessing $(\sim 10\%)$ and a trained classifier $(> 90\%)$, as seen in Fig \ref{fig:mnist_L2}, and Sibson mutual information consistently outperforms mutual information at almost all distortion levels. We also conduct further experiments to show that as the order of the Sibson mutual information increases, we obtain better privacy and lower adversary accuracies from Fig \ref{fig:mnist_L2}. Visualizations of the reconstructed digits can be seen in the supplementary materials.

\subsection{FERG data}
For the FERG data\cite{FERG} which consists of computer-generated faces of varying facial expressions, we pre-process the images into $50 \times 50$ gray-scale images and use them as inputs. We use two output labels, one for the regular task of predicting the expression, and the other for identifying the person's name. There are 7 different expressions, and 6 identities, thus our model's decoder component consists of two branches, one for the regular variable $Y$ and one for the private variable $C$. The distortion budget is the cross-entropy of the regular task label $Y$ with the output $\hat{Y}$, and subject to this budget the privatizer minimizes the Sibson mutual information for its parameters $\theta_p$. The decoder inference branch acts as an adversary that minimizes its cross-entropy for the private task over its parameters $\theta_a$. The distortion budget portion of the loss function is enforced as a penalty coefficient that increases with the number of iterations, same as equation (\ref{eq: minibatch L_p}).

\par The encoder consists of a neural network with 5 layers of 1024 hidden units with ReLU activations and $10\%$ dropout rate that maps the input gray-scale image into the parameters for a 512 dimensional isotropic Gaussian $Z$ distribution which is then averaged over a sample of 12 points. This is used as input to the decoder which outputs predictions for the two tasks via two branches, each consisting of fully connected neural networks of 3 layers of the same configuration of $(1024, 1024, 512)$ hidden units with $10\%$ dropout rate. The decoder outputs predicted probability vectors, one over the regular labels and one over the private labels. The optimization of the privatizer is subject to a budget on the cross-entropy loss for the regular task as a measure of the utility， while the adversary's objective is to minimize the cross-entropy of the private task with respect to the private branch parameters. Experiments for both Sibson mutual information and mutual information were conducted for distortion budgets ranging from $0.2$ to $1.8$, and the range was selected based on preliminary experiments. The adversary is trained for $k=20$ iterations for every iteration of training for the privatizer, and the entire model is trained over $200$ epochs with the Adam optimizer with a learning rate of $1e-3$ and a mini-batch size of $1000$.

\par
\begin{figure}[thpb]
      \centering
      \includegraphics[width=0.5 \textwidth]{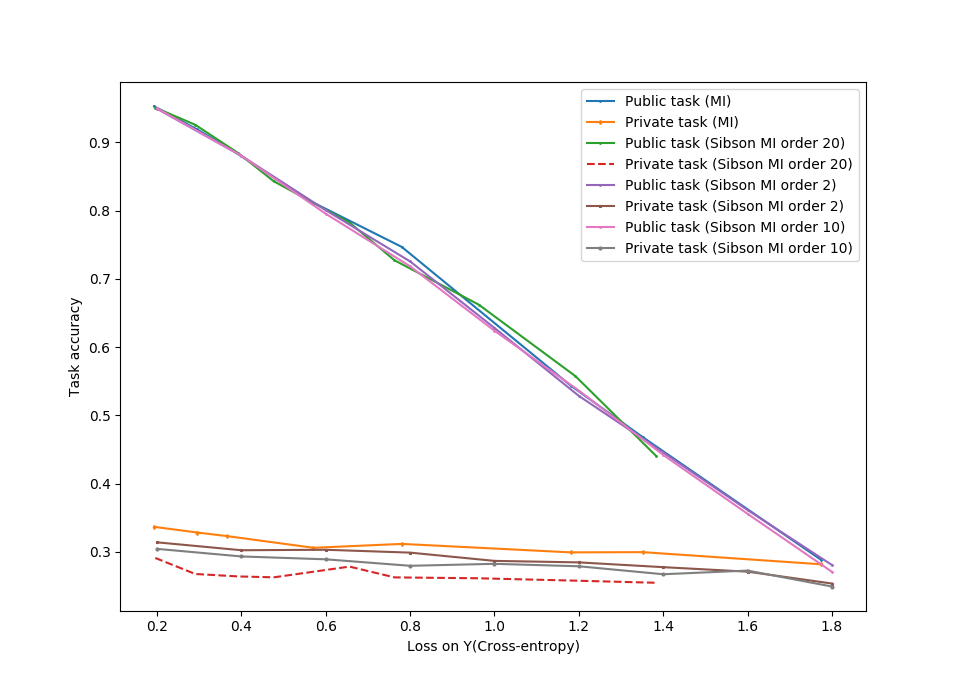}
      \caption{FERG model variant: accuracy rate vs distortion budget, when distortion is measured by log-loss of regular task}
      \label{fig:ferg_Dy_final_accs}
\end{figure}

\begin{figure}[thpb]
      \centering
      \includegraphics[width=0.4 \textwidth]{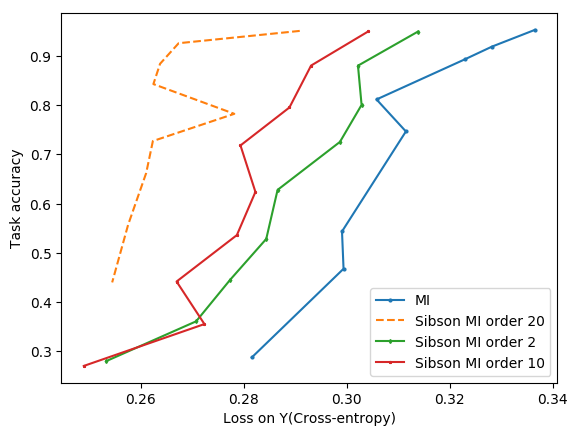}
      \caption{FERG model variant: regular task accuracy vs private task accuracy, when distortion is measured by log-loss of regular task}
      \label{fig:ferg_Dy_pairs}
\end{figure}

 The accuracy for regular and private tasks are plotted for both metrics over the experimental distortion budget calculated from the validation set. From Fig \ref{fig:ferg_Dy_final_accs} we can see that with different distortion budgets the model may leak little or substantial information with respect to the private variable, ranging from random guessing $(\sim 25\%)$ for the private task and little utility for the regular task $(\sim 45\%)$, to high probability $(> 90\%)$ of correctly guessing the regular label and $(\sim 30\%)$ for the private task. When using mutual information as the comparison metric, we find that the adversary performs on par in the public task but better in the private task across multiple distortion budgets, indicating worse privatization. We also plot the regular task versus private task accuracy for both metrics in Figure \ref{fig:ferg_Dy_pairs}, showing that Sibson mutual information provides more privacy than mutual information when holding the regular task accuracy fixed.


\begin{figure}[thpb]
      \centering
      \includegraphics[width=0.4 \textwidth]{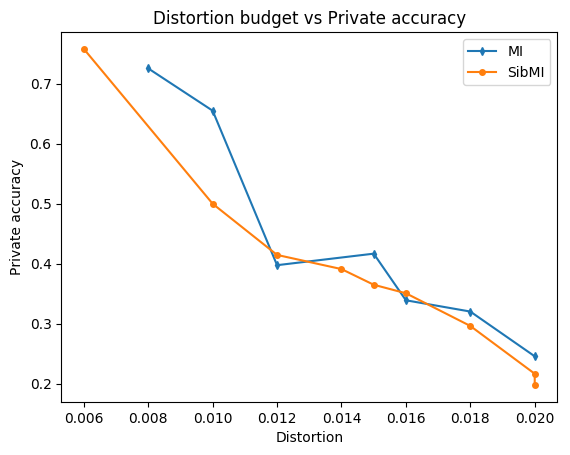}
      \caption{FERG data adversary accuracy rate vs distortion budget on X (lower is better), Sibson MI of order $20$}
      \label{fig:ferg_distx}
\end{figure}

We also consider one variant of this model that reconstructs the input, reducing it to the same as the previous experiment with MNIST data set. It uses a deconvolutional network using the same configuration as the MNIST data model with an extra fully connected layer to output the same dimensions as the input image. Then the privatizer is minimizing the Sibson mutual information subject to the reconstruction distortion budget from equation (\ref{eq: L2distortion}) while the adversary is trained to infer the private task. With this variant, the model was able to achieve various degrees of privacy-utility trade-off (23\% to 77\% adversary accuracy) based on a preset range of distortions as seen in Fig \ref{fig:ferg_distx}. It offers comparable or better privatization performance compared to mutual information again, as demonstrated by the lower adversary accuracy. Since this model variant aims to reconstruct $X$, we visualize the results in the supplementary materials as shown in Fig. \ref{fig:ferg_x_sample} for the original images. As the distortion budget increases, the model's reconstruction becomes increasingly blurry in Fig. \ref{fig:ferg_xhatvis_lowdist}, \ref{fig:ferg_xhatvis_highdist}. Another variant which is under development combines the reconstruction of the input with a regular task, and the overall distortion loss is a combination which the privatizer and adversary are trained to minimize within a budget. This model variant incentivizes the overall model to maintain a faithful reconstruction up to a degree and retain information useful towards accuracy in the regular task, while still minimizing the Sibson mutual information between the privatized representation and the private variable.

\section{Future work and conclusion} \label{sec: conclusion section}
For a theoretical data distribution scenario using affine transformations, we show that using maximal information leakage and Sibson mutual information as an optimization objective results in the same optimization problem as that of optimizing the MAP adversary accuracy, thus the optimal privatization mechanisms are equivalent. The experiments we conduct demonstrate that Sibson mutual information as a numerical proxy to maximal information leakage is an effective privacy metric for data-driven models to learn privacy mappings in order to reduce adversary performance. A possible future direction is to incorporate the decoded reconstruction as a generator for "natural" privatized data samples as determined by a discriminating network instead of measuring the reconstruction error by a set distortion metric. If the reconstruction from the privatizer can be used as a generated sample fed to a separate discriminator network, the discriminator can be trained to distinguish between real data samples and privatized data. The goal then, for the privatizer, is to learn a mapping that minimizes Sibson mutual information and has $50\%$ probability of being classified as a real data sample. We hope that this work will lead to wider usage of maximal information leakage in data disclosure systems and lead to stronger anonymization of user data.



\section{Appendix 1}
\subsection{Solution to optimization problem in section 4}
From the data setup and the optimization problem
\begin{flalign}
&\max_{(\beta_0, \beta_1) \in \mathcal{D}} \frac{\mu_0^{'} - \mu_1^{'}}{2\sigma}
\end{flalign}
we can rewrite as optimization over the parameters directly
\begin{flalign}
\max_{(\beta_0, \beta_1) \in \mathcal{D}} \beta_0 + \beta_1 \qquad s.t. \quad (1-\tilde{p})\beta_0^2 + \tilde{p}\beta_1^2 \le D,\\
\beta_0 + \beta_1 \le \mu_1 - \mu_0, \beta_0 \ge 0, \beta_1 \ge 0
\end{flalign}
The feasible region is defined by
\begin{flalign}
&(1-\tilde{p})\beta_0^2 + \tilde{p}\beta_1^2 \le D,\\
&\beta_0 \ge 0, \beta_1 \ge 0
\end{flalign}
when the equations $\beta_0 + \beta_1 = \mu_1 - \mu_0$ and $(1-\tilde{p})\beta_0^2 + \tilde{p}\beta_1^2 = D$ has no more than one solution, or
\begin{flalign}
    D \le \tilde{p}(1 - \tilde{p})(\mu_1 - \mu_0)^2
\end{flalign}
Under this situation, the distortion constraint is active, by applying the Karuhn-Kush-Tucker (KKT) conditions to 
\begin{flalign}
    \max_{(\beta_0, \beta_1) \in \mathcal{D}} (\beta_0 + \beta_1) + \gamma [(1-\tilde{p})\beta_0^2 + \tilde{p}\beta_1^2 - D]
\end{flalign}
we have the following equations to solve for:
\begin{flalign}
&1 + 2 \gamma^*(1 - \tilde{p})\beta_0^* = 0\\
&1 + 2 \gamma^* \tilde{p}\beta_1^* = 0\\
&(1-\tilde{p})\beta_0^{*2} + \tilde{p}\beta_1^{*2} - D = 0
\end{flalign}
Solving this set of equations for the variables $\beta_0^*, \beta_1^*, \gamma^*$ gives the optimal transformation parameters:
\begin{equation}
    \beta_0^{*2} = \ddfrac{\tilde{p}}{1-\tilde{p}} D, \qquad
    \beta_1^{*2} = \ddfrac{1-\tilde{p}}{\tilde{p}} D
\end{equation}
Otherwise the distortion budget constraint is not active and we may find a specific solution by setting the maximum distortion to
\begin{flalign}
&D = \tilde{p}(1 - \tilde{p})(\mu_1 - \mu_0)^2
\end{flalign}
then our expressions for the optimal parameters are:
\begin{flalign}
&\beta_0^* = \sqrt{\ddfrac{\tilde{p}}{1-\tilde{p}} D} = (\mu_1 - \mu_0)(1 - \tilde{p})\\
&\beta_1^* = \sqrt{\ddfrac{1-\tilde{p}}{\tilde{p}} D} = (\mu_1 - \mu_0)\tilde{p}
\end{flalign}
A general expression can be found by solving for the intersection of the distortion constraint $(1-\tilde{p})\beta_0^{*2} + \tilde{p}\beta_1^{*2} - D = 0$ and $\tilde{p}(1 - \tilde{p})(\mu_1 - \mu_0)^2 = D$ which in this case yields two solutions:
\begin{flalign}
&\beta_0^* = \tilde{p}(\mu_1 - \mu_0) \pm \sqrt{D + \tilde{p}(1 - \tilde{p})(\mu_1 - \mu_0)^2}\\
&\beta_1^* = \mu_1 - \mu_0 - \beta_0
\end{flalign}
All the points along the line segment with the endpoints of the two solutions for $(\beta_0^*, \beta_1^*)$ are optimal.

\subsection{Proof of Theorem 4}
\textbf{Proof:} Under the assumption that $\mu_0^{'} \le \mu_1^{'}$ we may compute the adversary's theoretical performance via a MAP decision rule:
\begin{flalign}
&Pr(\hat{C} = C)\\
&= \tilde{p}\displaystyle\int_{-\infty}^{z_0}P(Z|C=0)dz + (1 - \tilde{p})\displaystyle\int_{z_0}^{\infty}P(Z|C=1)dz\\
& z_0 = \frac{\sigma^2}{\mu_0^{'} - \mu_1^{'}}\log\big(\frac{1 - \tilde{p}}{\tilde{p}}\big) + \frac{\mu_0^{'} + \mu_1^{'}}{2}
\end{flalign}
where $z_0$ is derived by solving for $\tilde{p}P(Z|C=0) = (1 - \tilde{p})P(Z|C=1)$ under the MAP rule.
\begingroup
\allowdisplaybreaks
\begin{flalign}
&Pr(\hat{C} = C)\\
&= \tilde{p}\Big(1-Q\Big(\frac{z_0 - \mu_0^{'}}{\sigma}\Big)\Big) + (1 - \tilde{p})Q\Big(\frac{z_0 - \mu_1^{'}}{\sigma}\Big)\\
&= \tilde{p}Q\Big(-\frac{z_0^{'} - \mu_0^{'}}{\sigma}\Big) + (1 - \tilde{p})Q\Big(\frac{z_0^{'} - \mu_1^{'}}{\sigma}\Big)\\
&= \tilde{p}Q\Big(-\frac{\sigma}{\mu_0^{'} - \mu_1^{'}}\log(\frac{1 - \tilde{p}}{\tilde{p}}) + \frac{\mu_0^{'} - \mu_1^{'}}{2 \sigma}\Big)\\
&\quad + (1 - \tilde{p})Q\Big(\frac{\sigma}{\mu_0^{'} - \mu_1^{'}}\log(\frac{1 - \tilde{p}}{\tilde{p}}) + \frac{\mu_0^{'} - \mu_1^{'}}{2 \sigma}\Big) \nonumber\\
&= \tilde{p}Q\Big(\frac{1}{d}\log(\frac{1 - \tilde{p}}{\tilde{p}}) - \frac{d}{2}\Big) + (1 - \tilde{p})Q\Big(- \frac{1}{d}\log(\frac{1 - \tilde{p}}{\tilde{p}}) - \frac{d}{2}\Big)\\
&d = \frac{\mu_1^{'} - \mu_0^{'}}{\sigma} = \frac{\mu_1 - \mu_0 - (\beta_0 + \beta_1)}{\sigma}
\end{flalign}
\endgroup
We note that from
\begin{flalign}
&\frac{\partial (1-Q(x))}{\partial x} = -\frac{\partial Q(x)}{\partial x}\\
&= \frac{1}{\sqrt{2\pi}}\exp(-\frac{x^2}{2})
\end{flalign}
it is possible to compute the partial derivative w.r.t. $d$
\small
\begingroup
\allowdisplaybreaks
\begin{flalign}
&\frac{\partial Pr(\hat{C} = C)}{\partial d} \\
&=- \tilde{p} \frac{1}{\sqrt{2 \pi}} \exp\Big(-\big(-\frac{d}{2} + \frac{\log(\frac{1-\tilde{p}}{\tilde{p}})}{d}\big)^2/2\Big)\big(-\frac{1}{2} - \frac{\log(\frac{1-\tilde{p}}{\tilde{p}})}{d^2}\big)\\
&\quad - (1-\tilde{p}) \frac{1}{\sqrt{2 \pi}} \exp\Big(\big(\frac{d}{2} + \frac{\log(\frac{1-\tilde{p}}{\tilde{p}})}{d}\big)^2/2\Big)\big(-\frac{1}{2} + \frac{\log(\frac{1-\tilde{p}}{\tilde{p}})}{d^2}\big) \nonumber\\
&= -\tilde{p} \frac{1}{\sqrt{2 \pi}} \exp\Big(-\frac{d^2}{8} + \frac{\log(\frac{1 - \tilde{p}}{\tilde{p}})}{2} - \frac{\log(2\frac{1 - \tilde{p}}{\tilde{p}})}{2 d^2}\Big) \nonumber\\
&\quad \Big(-\frac{1}{2} - \frac{\log(\frac{1-\tilde{p}}{\tilde{p}})}{d^2}\Big)\\
&\quad - (1 - \tilde{p})\frac{1}{\sqrt{2 \pi}} \exp\Big(-\frac{d^2}{8} - \frac{\log(\frac{1 - \tilde{p}}{\tilde{p}})}{2} - \frac{\log(2\frac{1 - \tilde{p}}{\tilde{p}})}{2 d^2}\Big) \nonumber\\
&\quad \Big(-\frac{1}{2} + \frac{\log(\frac{1-\tilde{p}}{\tilde{p}})}{d^2}\Big) \nonumber\\
&= - \frac{1}{\sqrt{2 \pi}} \exp\Big(-\frac{d^2}{8} - \frac{\log(2\frac{1 - \tilde{p}}{\tilde{p}})}{2 d^2}\Big) \tilde{p} \exp\Big(\frac{\log(\frac{1 - \tilde{p}}{\tilde{p}})}{2}\Big)  \nonumber\\
&\quad \Big(-\frac{1}{2} - \frac{\log(\frac{1-\tilde{p}}{\tilde{p}})}{d^2}\Big)\\
&\quad - \frac{1}{\sqrt{2 \pi}} \exp\Big(-\frac{d^2}{8} - \frac{\log(2\frac{1 - \tilde{p}}{\tilde{p}})}{2 d^2}\Big) (1 - \tilde{p}) \exp\Big(\frac{\log(\frac{1 - \tilde{p}}{\tilde{p}})}{2}\Big) \nonumber\\
&\quad \Big(-\frac{1}{2} - \frac{\log(\frac{1-\tilde{p}}{\tilde{p}})}{d^2}\Big) \nonumber\\
&= - \frac{1}{\sqrt{2 \pi}} \exp\Big(-\frac{d^2}{8} - \frac{\log(2\frac{1 - \tilde{p}}{\tilde{p}})}{2 d^2}\Big) \Big[\tilde{p}\sqrt{\frac{1 - \tilde{p}}{\tilde{p}}}\big(-\frac{1}{2} - \frac{\log(\frac{1-\tilde{p}}{\tilde{p}})}{d^2}\big)\\
&\quad + (1-\tilde{p}) \sqrt{\frac{\tilde{p}}{1 - \tilde{p}}}\big(-\frac{1}{2} + \frac{\log(\frac{1-\tilde{p}}{\tilde{p}})}{d^2}\big)\Big] \nonumber\\
&= - \frac{1}{\sqrt{2 \pi}} \exp\Big(-\frac{d^2}{8} - \frac{\log(2\frac{1 - \tilde{p}}{\tilde{p}})}{2 d^2}\Big) \Big[\sqrt{\tilde{p}(1 - \tilde{p})}(-\frac{1}{2})\Big] > 0
\end{flalign}
\endgroup
\normalsize
Note that the objective is monotonically increasing in $d$, so directly minimizing the adversary's performance is equivalent to the optimization problem specified in subsection \ref{sec: affine optimization problem def}. The solution of this optimization is therefore the same as the solution in section A of Appendix 1.
\qedsymbol
\subsection{Derivation of monotonocity in Equation (\ref{eq: SibMI opt equals argmax d})}
\textbf{Proof:} With our approximation of Sibson mutual information, we have the optimization
\begin{flalign}
&\argmin_{(\beta_0, \beta_1) \in \mathcal{D}} \frac{\alpha}{\alpha - 1} \log(\tilde{p}^{1/\alpha}Q(\frac{1}{d \alpha}\log(\frac{1 - \tilde{p}}{\tilde{p}}) - \frac{d}{2}) + \\
&(1 - \tilde{p})Q(- \frac{1}{d \alpha}\log(\frac{1 - \tilde{p}}{\tilde{p}}) - \frac{d}{2}))\\
& = \argmin_{(\beta_0, \beta_1) \in \mathcal{D}} \frac{\alpha}{\alpha - 1} \log f(d, \alpha, \tilde{p}), \quad d = \frac{\mu_1^{'} - \mu_0^{'}}{\sigma}\\
\end{flalign}
Much like the previous section, here we will prove that optimization of the objective function above is also equivalent to equation (\ref{eq: SibMI opt equals argmax d}) by computing the derivative w.r.t. $d$:
\begingroup
\allowdisplaybreaks
\begin{flalign}
&\frac{\partial f(d, \alpha, \tilde{p})}{\partial d}\\
&=- \tilde{p}^{1/\alpha} \frac{1}{\sqrt{2 \pi}} \exp\Big(-\big(-\frac{d}{2} + \frac{\log(\frac{1-\tilde{p}}{\tilde{p}})}{d \alpha}\big)^2/2\Big) \\
&\quad \big(-\frac{1}{2} - \frac{\log(\frac{1-\tilde{p}}{\tilde{p}})}{d^2 \alpha}\big) \nonumber\\
&\quad - (1-\tilde{p})^{1/\alpha} \frac{1}{\sqrt{2 \pi}} \exp\Big(- \big(\frac{d}{2} + \frac{\log(\frac{1-\tilde{p}}{\tilde{p}})}{d \alpha}\big)^2/2\Big) \nonumber\\
&\quad \big(-\frac{1}{2} + \frac{\log(\frac{1-\tilde{p}}{\tilde{p}})}{d^2 \alpha}\big) \nonumber\\
&=- \frac{1}{\sqrt{2 \pi}} \exp\Big(-\frac{d^2}{8} - \frac{\log(2 \frac{1-\tilde{p}}{\tilde{p}})}{2 d^2 \alpha^2}\Big) \\
&\quad \Big[\tilde{p}^{1/\alpha} \exp\big(\frac{\log(\frac{1-\tilde{p}}{\tilde{p}})}{2 \alpha}\big) \big(-\frac{1}{2} - \frac{\log(\frac{1-\tilde{p}}{\tilde{p}})}{d^2 \alpha}\big) \nonumber\\
&\quad + (1-\tilde{p})^{1/\alpha} \exp\big(- \frac{\log(\frac{1-\tilde{p}}{\tilde{p}})}{2 \alpha}\big) \big(-\frac{1}{2} + \frac{\log(\frac{1-\tilde{p}}{\tilde{p}})}{d^2 \alpha}\big)\Big] \nonumber\\
&=- \frac{1}{\sqrt{2 \pi}} \exp\Big(-\frac{d^2}{8} - \frac{\log(2 \frac{1-\tilde{p}}{\tilde{p}})}{2 d^2 \alpha^2}\Big) \\
&\quad \Big[\tilde{p}^{1/\alpha} \exp\big(\frac{\log(\frac{1-\tilde{p}}{\tilde{p}})}{2 \alpha}\big) \big(-\frac{1}{2} - \frac{\log(\frac{1-\tilde{p}}{\tilde{p}})}{d^2 \alpha}\big) \nonumber\\
&\quad + (1-\tilde{p})^{1/\alpha} \exp\big(- \frac{\log(\frac{1-\tilde{p}}{\tilde{p}})}{2 \alpha}\big) \big(-\frac{1}{2} + \frac{\log(\frac{1-\tilde{p}}{\tilde{p}})}{d^2 \alpha}\big)\Big] \nonumber\\
&=- \frac{1}{\sqrt{2 \pi}} \exp\Big(-\frac{d^2}{8} - \frac{\log(2 \frac{1-\tilde{p}}{\tilde{p}})}{2 d^2 \alpha^2}\Big) \\
&\quad \Big[\tilde{p}^{1/\alpha} (\frac{1-\tilde{p}}{\tilde{p}})^{\frac{1}{2 \alpha}} \big(-\frac{1}{2} - \frac{\log(\frac{1-\tilde{p}}{\tilde{p}})}{d^2 \alpha}\big) \nonumber\\
&\quad + (1-\tilde{p})^{1/\alpha} (\frac{\tilde{p}}{1-\tilde{p}})^{\frac{1}{2 \alpha}} \big(-\frac{1}{2} + \frac{\log(\frac{1-\tilde{p}}{\tilde{p}})}{d^2 \alpha}\big)\Big] \nonumber\\
&= - \frac{1}{\sqrt{2 \pi}} \exp\Big(-\frac{d^2}{8} - \frac{\log(2 \frac{1-\tilde{p}}{\tilde{p}})}{2 d^2 \alpha^2}\Big) \big[\tilde{p}(1-\tilde{p})\big]^{\frac{1}{2 \alpha}} \\
&\quad \Big(-\frac{1}{2} - \frac{\log(\frac{1-\tilde{p}}{\tilde{p}})}{d^2 \alpha} + \big(-\frac{1}{2} + \frac{\log(\frac{1-\tilde{p}}{\tilde{p}})}{d^2 \alpha}\big) \Big) > 0 \nonumber
\end{flalign}
\endgroup
Thus the optimization objective is monotonically increasing in $d$, and is equivalent to equation (\ref{eq: SibMI opt equals argmax d}).

\section*{ACKNOWLEDGMENT}
The authors would like to thank Vincent Y.F. Tan for his insights and advice in the course of developing this work.


\bibliographystyle{IEEEtran}
\bibliography{Bibliography}

\begin{thebibliography}{10}
\providecommand{\url}[1]{#1}
\csname url@rmstyle\endcsname
\providecommand{\newblock}{\relax}
\providecommand{\bibinfo}[2]{#2}
\providecommand\BIBentrySTDinterwordspacing{\spaceskip=0pt\relax}
\providecommand\BIBentryALTinterwordstretchfactor{4}
\providecommand\BIBentryALTinterwordspacing{\spaceskip=\fontdimen2\font plus
\BIBentryALTinterwordstretchfactor\fontdimen3\font minus
  \fontdimen4\font\relax}
\providecommand\BIBforeignlanguage[2]{{%
\expandafter\ifx\csname l@#1\endcsname\relax
\typeout{** WARNING: IEEEtran.bst: No hyphenation pattern has been}%
\typeout{** loaded for the language `#1'. Using the pattern for}%
\typeout{** the default language instead.}%
\else
\language=\csname l@#1\endcsname
\fi
#2}}

\bibitem{SweeneySimpleDemographics}
\BIBentryALTinterwordspacing
L.~Sweeney, ``Simple demographics often identify people uniquely,'' 2000.
  [Online]. Available:
  \url{http://dataprivacylab.org/projects/identifiability/}
\BIBentrySTDinterwordspacing

\bibitem{NarayananRobustDeanon}
\BIBentryALTinterwordspacing
A.~Narayanan and V.~Shmatikov, ``Robust de-anonymization of large sparse
  datasets,'' in \emph{Proceedings of the 2008 IEEE Symposium on Security and
  Privacy}, ser. SP '08.\hskip 1em plus 0.5em minus 0.4em\relax Washington, DC,
  USA: IEEE Computer Society, 2008, pp. 111--125. [Online]. Available:
  \url{https://doi.org/10.1109/SP.2008.33}
\BIBentrySTDinterwordspacing

\bibitem{DBLP:conf/fossacs/Smith09}
\BIBentryALTinterwordspacing
G.~Smith, ``On the foundations of quantitative information flow,'' in
  \emph{Foundations of Software Science and Computational Structures, 12th
  International Conference, {FOSSACS} 2009, Held as Part of the Joint European
  Conferences on Theory and Practice of Software, {ETAPS} 2009, York, UK, March
  22-29, 2009. Proceedings}, ser. Lecture Notes in Computer Science,
  L.~de~Alfaro, Ed., vol. 5504.\hskip 1em plus 0.5em minus 0.4em\relax
  Springer, 2009, pp. 288--302. [Online]. Available:
  \url{https://doi.org/10.1007/978-3-642-00596-1\_21}
\BIBentrySTDinterwordspacing

\bibitem{DBLP:journals/entcs/BraunCP09}
\BIBentryALTinterwordspacing
C.~Braun, K.~Chatzikokolakis, and C.~Palamidessi, ``Quantitative notions of
  leakage for one-try attacks,'' \emph{Electr. Notes Theor. Comput. Sci.}, vol.
  249, pp. 75--91, 2009. [Online]. Available:
  \url{https://doi.org/10.1016/j.entcs.2009.07.085}
\BIBentrySTDinterwordspacing

\bibitem{DBLP:conf/csfw/BartheK11}
\BIBentryALTinterwordspacing
G.~Barthe and B.~K{\"{o}}pf, ``Information-theoretic bounds for differentially
  private mechanisms,'' in \emph{Proceedings of the 24th {IEEE} Computer
  Security Foundations Symposium, {CSF} 2011, Cernay-la-Ville, France, 27-29
  June, 2011}.\hskip 1em plus 0.5em minus 0.4em\relax {IEEE} Computer Society,
  2011, pp. 191--204. [Online]. Available:
  \url{https://doi.org/10.1109/CSF.2011.20}
\BIBentrySTDinterwordspacing

\bibitem{PPANpaper}
\BIBentryALTinterwordspacing
A.~Tripathy, Y.~Wang, and P.~Ishwar, ``Privacy-preserving adversarial
  networks,'' \emph{CoRR}, vol. abs/1712.07008, 2017. [Online]. Available:
  \url{http://arxiv.org/abs/1712.07008}
\BIBentrySTDinterwordspacing

\bibitem{MinimaxFilter}
\BIBentryALTinterwordspacing
J.~Hamm, ``Minimax filter: Learning to preserve privacy from inference
  attacks,'' \emph{CoRR}, vol. abs/1610.03577, 2016. [Online]. Available:
  \url{http://arxiv.org/abs/1610.03577}
\BIBentrySTDinterwordspacing

\bibitem{DworkFoundationofDiffPriv}
\BIBentryALTinterwordspacing
C.~Dwork, ``A firm foundation for private data analysis,'' \emph{Commun. ACM},
  vol.~54, no.~1, pp. 86--95, Jan. 2011. [Online]. Available:
  \url{http://doi.acm.org/10.1145/1866739.1866758}
\BIBentrySTDinterwordspacing

\bibitem{DeepLearningwDiffPriv}
\BIBentryALTinterwordspacing
M.~Abadi, A.~Chu, I.~J. Goodfellow, H.~B. McMahan, I.~Mironov, K.~Talwar, and
  L.~Zhang, ``Deep learning with differential privacy,'' \emph{CoRR}, vol.
  abs/1607.00133, 2016. [Online]. Available:
  \url{http://arxiv.org/abs/1607.00133}
\BIBentrySTDinterwordspacing

\bibitem{DifPrivEmpRiskMin}
\BIBentryALTinterwordspacing
K.~Chaudhuri, C.~Monteleoni, and A.~D. Sarwate, ``Differentially private
  empirical risk minimization,'' \emph{J. Mach. Learn. Res.}, vol.~12, pp.
  1069--1109, July 2011. [Online]. Available:
  \url{http://dl.acm.org/citation.cfm?id=1953048.2021036}
\BIBentrySTDinterwordspacing

\bibitem{SGDwithPrivUpdates}
\BIBentryALTinterwordspacing
S.~Song, K.~Chaudhuri, and A.~D. Sarwate, ``Stochastic gradient descent with
  differentially private updates,'' in \emph{{IEEE} Global Conference on Signal
  and Information Processing, GlobalSIP 2013, Austin, TX, USA, December 3-5,
  2013}.\hskip 1em plus 0.5em minus 0.4em\relax {IEEE}, 2013, pp. 245--248.
  [Online]. Available: \url{https://doi.org/10.1109/GlobalSIP.2013.6736861}
\BIBentrySTDinterwordspacing

\bibitem{PrivfromMultipartyData}
\BIBentryALTinterwordspacing
J.~Hamm, P.~Cao, and M.~Belkin, ``Learning privately from multiparty data,''
  \emph{CoRR}, vol. abs/1602.03552, 2016. [Online]. Available:
  \url{http://arxiv.org/abs/1602.03552}
\BIBentrySTDinterwordspacing

\bibitem{Shokri:2015:PDL:2810103.2813687}
\BIBentryALTinterwordspacing
R.~Shokri and V.~Shmatikov, ``Privacy-preserving deep learning,'' in
  \emph{Proceedings of the 22Nd ACM SIGSAC Conference on Computer and
  Communications Security}, ser. CCS '15.\hskip 1em plus 0.5em minus
  0.4em\relax New York, NY, USA: ACM, 2015, pp. 1310--1321. [Online].
  Available: \url{http://doi.acm.org/10.1145/2810103.2813687}
\BIBentrySTDinterwordspacing

\bibitem{VIME}
\BIBentryALTinterwordspacing
R.~Houthooft, X.~Chen, Y.~Duan, J.~Schulman, F.~D. Turck, and P.~Abbeel,
  ``{VIME:} variational information maximizing exploration,'' in \emph{Advances
  in Neural Information Processing Systems 29: Annual Conference on Neural
  Information Processing Systems 2016, December 5-10, 2016, Barcelona, Spain},
  D.~D. Lee, M.~Sugiyama, U.~von Luxburg, I.~Guyon, and R.~Garnett, Eds., 2016,
  pp. 1109--1117. [Online]. Available:
  \url{http://papers.nips.cc/paper/6591-vime-variational-information-maximizing-exploration}
\BIBentrySTDinterwordspacing

\bibitem{GoodfellowGAN}
\BIBentryALTinterwordspacing
I.~J. Goodfellow, J.~Pouget{-}Abadie, M.~Mirza, B.~Xu, D.~Warde{-}Farley,
  S.~Ozair, A.~C. Courville, and Y.~Bengio, ``Generative adversarial nets,'' in
  \emph{Advances in Neural Information Processing Systems 27: Annual Conference
  on Neural Information Processing Systems 2014, December 8-13 2014, Montreal,
  Quebec, Canada}, Z.~Ghahramani, M.~Welling, C.~Cortes, N.~D. Lawrence, and
  K.~Q. Weinberger, Eds., 2014, pp. 2672--2680. [Online]. Available:
  \url{http://papers.nips.cc/paper/5423-generative-adversarial-nets}
\BIBentrySTDinterwordspacing

\bibitem{ImprovedtechniquesGANs}
\BIBentryALTinterwordspacing
T.~Salimans, I.~J. Goodfellow, W.~Zaremba, V.~Cheung, A.~Radford, and X.~Chen,
  ``Improved techniques for training gans,'' \emph{CoRR}, vol. abs/1606.03498,
  2016. [Online]. Available: \url{http://arxiv.org/abs/1606.03498}
\BIBentrySTDinterwordspacing

\bibitem{InfoGAN}
\BIBentryALTinterwordspacing
X.~Chen, Y.~Duan, R.~Houthooft, J.~Schulman, I.~Sutskever, and P.~Abbeel,
  ``Infogan: Interpretable representation learning by information maximizing
  generative adversarial nets,'' in \emph{Advances in Neural Information
  Processing Systems 29: Annual Conference on Neural Information Processing
  Systems 2016, December 5-10, 2016, Barcelona, Spain}, D.~D. Lee, M.~Sugiyama,
  U.~von Luxburg, I.~Guyon, and R.~Garnett, Eds., 2016, pp. 2172--2180.
  [Online]. Available:
  \url{http://papers.nips.cc/paper/6399-infogan-interpretable-representation-learning-by-\\information-maximizing-generative-adversarial-nets}
\BIBentrySTDinterwordspacing

\bibitem{CycleGAN}
\BIBentryALTinterwordspacing
J.~Zhu, T.~Park, P.~Isola, and A.~A. Efros, ``Unpaired image-to-image
  translation using cycle-consistent adversarial networks,'' \emph{CoRR}, vol.
  abs/1703.10593, 2017. [Online]. Available:
  \url{http://arxiv.org/abs/1703.10593}
\BIBentrySTDinterwordspacing

\bibitem{MakhzaniAdvAutoencoders}
\BIBentryALTinterwordspacing
A.~Makhzani, J.~Shlens, N.~Jaitly, and I.~J. Goodfellow, ``Adversarial
  autoencoders,'' \emph{CoRR}, vol. abs/1511.05644, 2015. [Online]. Available:
  \url{http://arxiv.org/abs/1511.05644}
\BIBentrySTDinterwordspacing

\bibitem{CGAPpaper}
\BIBentryALTinterwordspacing
C.~Huang, P.~Kairouz, X.~Chen, L.~Sankar, and R.~Rajagopal, ``Context-aware
  generative adversarial privacy,'' \emph{CoRR}, vol. abs/1710.09549, 2017.
  [Online]. Available: \url{http://arxiv.org/abs/1710.09549}
\BIBentrySTDinterwordspacing

\bibitem{DBLP:journals/information/AsoodehDAL16}
\BIBentryALTinterwordspacing
S.~Asoodeh, M.~Diaz, F.~Alajaji, and T.~Linder, ``Information extraction under
  privacy constraints,'' \emph{Information}, vol.~7, no.~1, p.~15, 2016.
  [Online]. Available: \url{https://doi.org/10.3390/info7010015}
\BIBentrySTDinterwordspacing

\bibitem{DBLP:conf/cwit/AsoodehAL15}
\BIBentryALTinterwordspacing
S.~Asoodeh, F.~Alajaji, and T.~Linder, ``On maximal correlation, mutual
  information and data privacy,'' in \emph{14th {IEEE} Canadian Workshop on
  Information Theory, {CWIT} 2015, St. John's, NL, Canada, July 6-9,
  2015}.\hskip 1em plus 0.5em minus 0.4em\relax {IEEE}, 2015, pp. 27--31.
  [Online]. Available: \url{https://doi.org/10.1109/CWIT.2015.7255145}
\BIBentrySTDinterwordspacing

\bibitem{IssaWagnerOperationalMeasure}
\BIBentryALTinterwordspacing
I.~Issa, S.~Kamath, and A.~B. Wagner, ``An operational measure of information
  leakage,'' in \emph{2016 Annual Conference on Information Science and
  Systems, {CISS} 2016, Princeton, NJ, USA, March 16-18, 2016}.\hskip 1em plus
  0.5em minus 0.4em\relax {IEEE}, 2016, pp. 234--239. [Online]. Available:
  \url{https://doi.org/10.1109/CISS.2016.7460507}
\BIBentrySTDinterwordspacing

\bibitem{AlvimGeneralizedGainFunctions}
\BIBentryALTinterwordspacing
M.~S. Alvim, K.~Chatzikokolakis, C.~Palamidessi, and G.~Smith, ``Measuring
  information leakage using generalized gain functions,'' in \emph{Proceedings
  of the 2012 IEEE 25th Computer Security Foundations Symposium}, ser. CSF
  '12.\hskip 1em plus 0.5em minus 0.4em\relax Washington, DC, USA: IEEE
  Computer Society, 2012, pp. 265--279. [Online]. Available:
  \url{http://dx.doi.org/10.1109/CSF.2012.26}
\BIBentrySTDinterwordspacing

\bibitem{Axiomsforleakage}
\BIBentryALTinterwordspacing
M.~S. Alvim, K.~Chatzikokolakis, A.~McIver, C.~Morgan, C.~Palamidessi, and
  G.~Smith, ``Axioms for information leakage,'' in \emph{{IEEE} 29th Computer
  Security Foundations Symposium, {CSF} 2016, Lisbon, Portugal, June 27 - July
  1, 2016}.\hskip 1em plus 0.5em minus 0.4em\relax {IEEE} Computer Society,
  2016, pp. 77--92. [Online]. Available:
  \url{https://doi.org/10.1109/CSF.2016.13}
\BIBentrySTDinterwordspacing

\bibitem{Csiszar95GenCutoffandRenyiInfoMeasures}
\BIBentryALTinterwordspacing
I.~Csisz{\'{a}}r, ``Generalized cutoff rates and renyi's information
  measures,'' \emph{{IEEE} Trans. Information Theory}, vol.~41, no.~1, pp.
  26--34, 1995. [Online]. Available: \url{https://doi.org/10.1109/18.370121}
\BIBentrySTDinterwordspacing

\bibitem{Ben-BassatR78RenyiEntropy}
\BIBentryALTinterwordspacing
M.~Ben{-}Bassat and J.~Raviv, ``Renyi's entropy and the probability of error,''
  \emph{{IEEE} Trans. Information Theory}, vol.~24, no.~3, pp. 324--331, 1978.
  [Online]. Available: \url{https://doi.org/10.1109/TIT.1978.1055890}
\BIBentrySTDinterwordspacing

\bibitem{VerdualphaMI}
\BIBentryALTinterwordspacing
S.~Verd{\'{u}}, ``{\(\alpha\)}-mutual information,'' in \emph{2015 Information
  Theory and Applications Workshop, {ITA} 2015, San Diego, CA, USA, February
  1-6, 2015}.\hskip 1em plus 0.5em minus 0.4em\relax {IEEE}, 2015, pp. 1--6.
  [Online]. Available: \url{https://doi.org/10.1109/ITA.2015.7308959}
\BIBentrySTDinterwordspacing

\bibitem{IssaW17OperationalDefinitions}
\BIBentryALTinterwordspacing
I.~Issa and A.~B. Wagner, ``Operational definitions for some common information
  leakage metrics,'' in \emph{2017 {IEEE} International Symposium on
  Information Theory, {ISIT} 2017, Aachen, Germany, June 25-30, 2017}.\hskip
  1em plus 0.5em minus 0.4em\relax {IEEE}, 2017, pp. 769--773. [Online].
  Available: \url{https://doi.org/10.1109/ISIT.2017.8006632}
\BIBentrySTDinterwordspacing

\bibitem{NormalDistributions}
\BIBentryALTinterwordspacing
E.~Weisstein, ``Normal distribution,'' 2002. [Online]. Available:
  \url{http://mathworld.wolfram.com/NormalDistribution.html}
\BIBentrySTDinterwordspacing

\bibitem{KingmaWellingAutoencodingVB}
\BIBentryALTinterwordspacing
D.~P. Kingma and M.~Welling, ``Auto-encoding variational bayes,'' \emph{CoRR},
  vol. abs/1312.6114, 2013. [Online]. Available:
  \url{http://arxiv.org/abs/1312.6114}
\BIBentrySTDinterwordspacing

\bibitem{adam}
\BIBentryALTinterwordspacing
D.~P. Kingma and J.~Ba, ``Adam: {A} method for stochastic optimization,''
  \emph{CoRR}, vol. abs/1412.6980, 2014. [Online]. Available:
  \url{http://arxiv.org/abs/1412.6980}
\BIBentrySTDinterwordspacing

\bibitem{FERG}
D.~Aneja, A.~Colburn, G.~Faigin, L.~Shapiro, and B.~Mones, ``Modeling stylized
  character expressions via deep learning,'' in \emph{Asian Conference on
  Computer Vision}.\hskip 1em plus 0.5em minus 0.4em\relax Springer, 2016, pp.
  136--153.

\end{thebibliography}





\vfill


\section*{Supplementary Materials}

\subsection*{Visualizations of MNIST data reconstructions} \label{sec: MNIST visualizations}
\par Below in Figures \ref{fig:mnist_x_samples}-\ref{fig:mnist_xhatvis_0.08dist} are shown visualizations of the MNIST digits, from the original image to reconstructions with increasing distortion budgets. They are outputs from the model under Sibson mutual information as the private metric.

\begin{figure}[thpb]
      \centering
      \includegraphics[width=0.4 \textwidth]{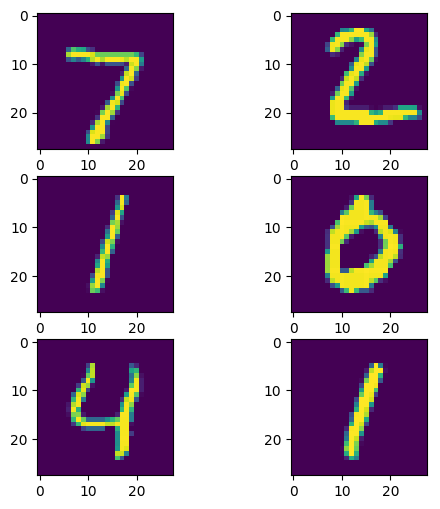}
      \caption{MNIST data samples}
      \label{fig:mnist_x_samples}
\end{figure}
\begin{figure}[thpb]
      \centering
      \includegraphics[width=0.4 \textwidth]{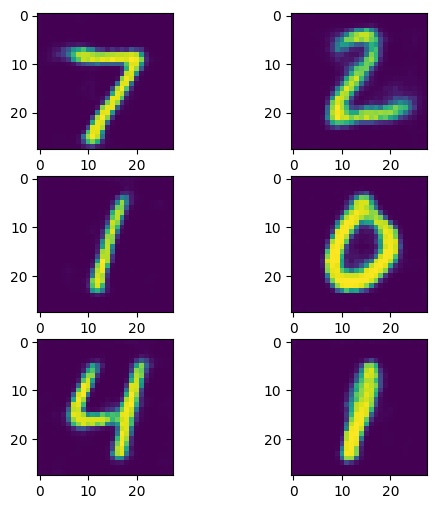}
      \caption{MNIST data reconstructions with a 0.02 distortion budget on X (Sibson MI)}
      \label{fig:mnist_xhatvis_0.02dist}
\end{figure}
\begin{figure}[thpb]
      \centering
      \includegraphics[width=0.4 \textwidth]{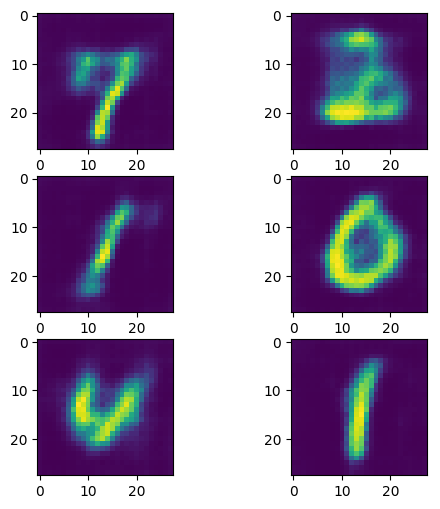}
      \caption{MNIST data reconstructions with a 0.04 distortion budget on X(Sibson MI)}
      \label{fig:mnist_xhatvis_0.04dist}
\end{figure}
\begin{figure}[thpb]
      \centering
      \includegraphics[width=0.4 \textwidth]{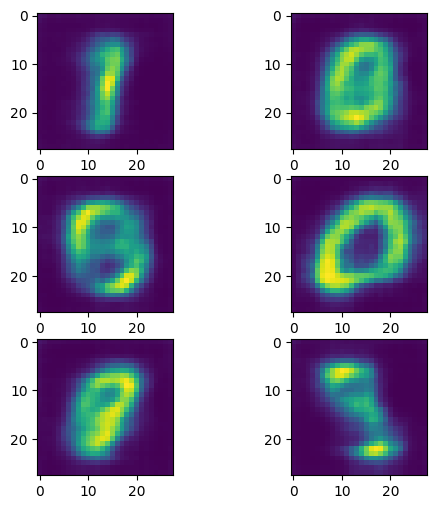}
      \caption{MNIST data reconstructions with a 0.06 distortion budget on X(Sibson MI)}
      \label{fig:mnist_xhatvis_0.06dist}
\end{figure}
\begin{figure}[thpb]
      \centering
      \includegraphics[width=0.4 \textwidth]{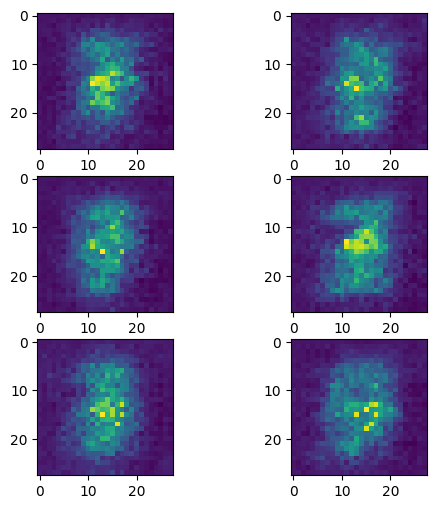}
      \caption{MNIST data reconstructions with a 0.08 distortion budget on X(Sibson MI)}
      \label{fig:mnist_xhatvis_0.08dist}
\end{figure}

\subsection*{Visualizations of FERG data reconstructions} \label{sec: FERG visualizations}
\par Below in Figures \ref{fig:ferg_x_sample}-\ref{fig:ferg_xhatvis_highdist} are shown visualizations of the FERG reconstructions, from the original image to reconstructions with low and high distortion budgets. They are outputs from the model under Sibson mutual information of order $20$ as the private metric.

\begin{figure}[thpb]
      \centering
      \includegraphics[width=0.4 \textwidth]{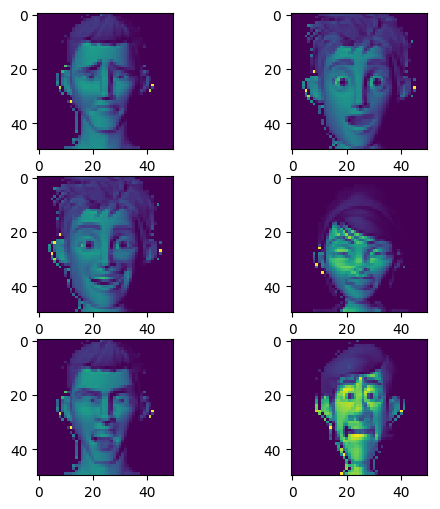}
      \caption{FERG data samples}
      \label{fig:ferg_x_sample}
\end{figure}
\begin{figure}[thpb]
      \centering
      \includegraphics[width=0.4 \textwidth]{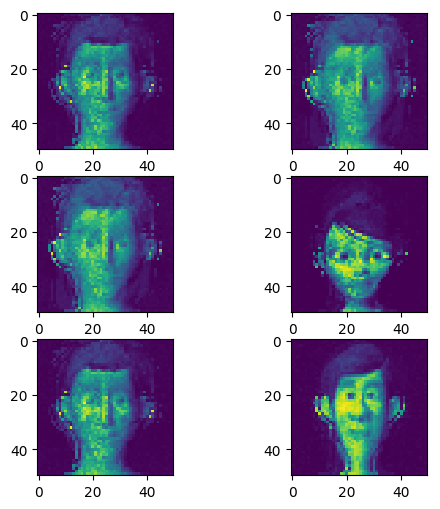}
      \caption{FERG data reconstructions with a 0.006 distortion budget on X}
      \label{fig:ferg_xhatvis_lowdist}
\end{figure}
\begin{figure}[thpb]
      \centering
      \includegraphics[width=0.4 \textwidth]{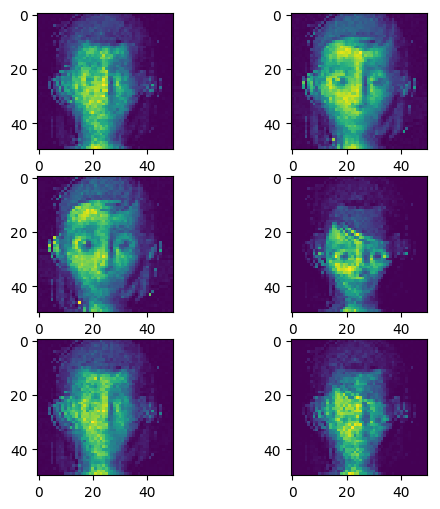}
      \caption{FERG data reconstructions with a 0.01 distortion budget on X}
      \label{fig:ferg_xhatvis_highdist}
\end{figure}

\end{document}